\documentclass[journal]{IEEEtran}

\usepackage{cite}

\usepackage{makeidx,epsfig}
\usepackage{setspace,graphicx,multirow}

\usepackage{amsfonts,latexsym,amssymb,amsthm}

\usepackage{graphicx}
\usepackage{epstopdf}

\usepackage{caption3} 
\DeclareCaptionOption{parskip}[]{} 
\usepackage[small]{caption}

\usepackage{subcaption}

\usepackage{array}

\usepackage[cmex10]{amsmath}

\usepackage{url}

\usepackage{multirow}
\usepackage{hhline}

\usepackage{amsmath}
\usepackage{mathtools}

\usepackage{algorithm}
\usepackage[]{algpseudocode}
\usepackage{varwidth}

\makeatletter
\def\BState{\State\hskip-\ALG@thistlm}
\makeatother

\usepackage{algcompatible}

\usepackage{fixltx2e}

\newcommand*\diff{\mathop{}\!\mathrm{d}}

\usepackage{enumitem}

\makeatletter
  \newcommand\tinyv{\@setfontsize\tinyv{7pt}{9}}
\makeatother

\usepackage{authblk}

\usepackage{textcomp}

\usepackage{color}

\renewcommand\Authands{ and }

\begin{document}
\bibliographystyle{IEEEtran}
\bstctlcite{IEEEexample:BSTcontrol}

\title{Network-Connected UAV: 3D System Modeling and Coverage Performance Analysis}

\author{Jiangbin~Lyu,~\textit{Member,~IEEE},
        and~Rui~Zhang,~\textit{Fellow,~IEEE}%
\thanks{J. Lyu is with School of Information Science and Engineering, and Key Laboratory of Underwater Acoustic Communication and Marine Information Technology, Xiamen University, China 361005 (e-mail: ljb@xmu.edu.cn).}
\thanks{R. Zhang is with the Department of Electrical and Computer Engineering, National University of Singapore, Singapore 117583 (email: elezhang@nus.edu.sg).}
}

\maketitle


\begin{abstract}
With growing popularity, unmanned aerial vehicles (UAVs) are pivotally extending conventional terrestrial Internet of Things (IoT) into the sky. To enable high-performance two-way communications of UAVs with their ground pilots/users, \textit{cellular network-connected UAV} has drawn significant interests recently. Among others, an important issue is whether the existing cellular network, designed mainly for terrestrial users, is also able to effectively cover the new UAV users in the three-dimensional (3D) space for both uplink and downlink communications. Such 3D coverage analysis is challenging due to the unique air-ground channel characteristics, the resulted interference issue with terrestrial communication, and the non-uniform 3D antenna gain pattern of ground base station (GBS) in practice. Particularly, high-altitude UAV often possesses a high probability of line-of-sight (LoS) channels with a large number of GBSs, while their random binary (LoS/Non-LoS) channel states and (on/off) activities give rise to exponentially large number of discrete UAV-GBS association/interference states, rendering coverage analysis more difficult. This paper presents a new 3D system model to incorporate UAV users and proposes an analytical framework to characterize their uplink/downlink 3D coverage performance. To tackle the above exponential complexity, we introduce a \textit{generalized Poisson multinomial (GPM)} distribution to model the discrete interference states, and a novel \textit{lattice approximation (LA)} technique to approximate the non-lattice GPM variable and obtain the interference distribution efficiently with high accuracy. The 3D coverage analysis is validated by extensive numerical results, which also show effects of key system parameters such as cell loading factor, GBS antenna downtilt, UAV altitude and antenna beamwidth.
\end{abstract}

\section{Introduction}

%
%
%
%
%
%

With enhanced functionality and reducing cost, unmanned aerial vehicles (UAVs), or so-called drones, have found fast-growing applications over recent years in the civilian domain such as for cargo delivery, precise agriculture, aerial imaging, search and rescue, etc.
In particular, UAV can be employed as aerial communication platform\cite{ZengUAVmag} to provide wireless connectivity for the ground users and Internet of Things (IoT) devices\cite{FengWei} when traditional terrestrial networks are unavailable, insufficient or costly to deploy.
Typical applications include UAV-aided communication offloading for temporary hotspot regions\cite{CyclicalLyu,UAVGBSTWC,LAPlosProbability,ZhangWeiSmallCellJSAC, QingqingTrajectory,Jingwei1,Dhillon2017}; mobile data relaying between distant ground users\cite{ZengMobileRelay}; and efficient information dissemination or data collection in IoT and sensor networks\cite{ZhanChengDataCollection}\cite{YangDingChengTVT}, etc.
On the other hand, UAVs can be integrated into the existing and future cellular networks as new aerial user equipments (UEs) to enable their two-way communications with the terrestrial users efficiently, thus extending the IoT to the sky, known as the \textit{Internet of Drones (IoD)}\cite{IoD}.
To achieve the cellular-enabled IoD, 
it is of paramount importance to ensure that all UAVs can operate safely and reliably, even in harsh environments. This calls for ultra-reliable, low-latency, and secure communication links between the ground base stations (GBSs) and the UAV for supporting the critical control and non-payload communications (CNPC). Moreover, in many applications such as video streaming, surveillance and aerial imaging, UAVs generally require high-capacity data communication links with the GBSs so as to timely send the payload data (such as high-quality images and videos) to the end users.

To enable high-performance two-way communications between UAVs and ground users, the existing 4G (fourth-generation) LTE (Long Term Evolution) or forthcoming 5G (fifth-generation) cellular networks can be leveraged thanks to their almost ubiquitous accessibility and superior performance.
As a result, \textit{network-connected UAV communications} have drawn significant interests recently (see e.g. \cite{UAVinterferenceMagazine,SkyIsNotLimit,CellularConnectedUAVmag,CellularUAVshuowen} and the references therein).
In fact, the 3rd Generation Partnership Project (3GPP) had launched a new work item to investigate the various issues and their solutions for UAV communications in the current LTE network\cite{3GPPworkItemUAV}.
Moreover, increasingly more field trials have been conducted on using terrestrial cellular networks to provide wireless connectivity for UAVs \cite{QualCommDroneReport}\cite{LinXingqinSimulations}.

Among others, 
one critical issue to address for cellular-enabled UAV communications is whether the cellular network is able to provide reliable three-dimensional (3D) coverage for the UAVs at various altitude in both the uplink and downlink communications\footnote{We follow the convention to use ``downlink" to refer to the communication from GBS to UE and ``uplink" to that in the reverse direction, although UAVs usually have much higher altitude than GBSs in practice.}.
To achieve reliable UAV-GBS communications, the uplink or downlink signal-to-noise ratio (SNR), with co-channel interference treated as additional noise, needs to be no smaller than a predefined threshold; otherwise, an \textit{outage} will occur. 
For the UAV's uplink, it may transmit multimedia data to the GBS, which requires high data rate and correspondingly high SNR threshold.
On the other hand, the UAV's downlink communication needs to support the critical CNPC data from its associated GBS, which is typically of lower data rate but requires higher reliability (lower outage probability) compared to the uplink.
The \textit{3D coverage probability} of the UAV uplink/downlink communication is thus defined as the corresponding average non-outage probability of a UAV uniformly located in a given 3D space.

The 3D coverage performance analysis for aerial users is considerably different from its two-dimensional (2D) counterpart for ground UEs in the traditional cellular network\cite{AndrewsCellular}, due to their distinct channel characteristics and resulted interference effects. Specifically, for ground UEs, their channels with GBSs usually exhibit severe pathloss and prominent small-scale fading due to the rich scattering environment especially in urban areas.
In contrast, for UAVs in the sky far above the GBSs, their communication signals are more likely to propagate through free space with few obstacles, and hence line-of-sight (LoS) links usually exist with a high probability, which increases with the UAV altitude in general \cite{3GPPevaluationAssumption}.
Although LoS channels entail more reliable communication between the UAV and its serving GBS as compared to the terrestrial UEs, they also cause 
more severe uplink/downlink interference to/from a larger number of non-associated GBSs.
This thus calls for effective interference management techniques such as multi-cell coordinated channel assignment and transmission \cite{WeidongUAVuplink}, advanced antenna beamforming techniques at the UAV\cite{LiuLiangUAVuplink} and GBS\cite{MassiveMIMOuav}, interference-aware UAV path planning\cite{CellularUAVshuowen,WalidDeepLearning}, etc.
Besides the LoS dominant air-ground channel, the 3D GBS antenna pattern also has a significant effect on the UAV's coverage performance.
The commonly adopted GBS antenna pattern in 3GPP \cite{3GPPsimplifiedElevation} or existing literature \cite{AndrewsCellular}\cite{PollinGC2017} is usually simplified in the vertical domain, e.g., by specifying only fixed gains for the antenna mainlobe and sidelobes, respectively. 
In practice, however, the GBS antenna is usually tilted downward to support ground UEs \cite{QualCommDroneReport}, and hence likely to communicate with aerial UEs in its sidelobes only.
Therefore, a more practically refined model for the GBS antenna pattern is needed to characterize the 3D coverage performance of the UAV accurately, especially for the case with non-uniform sidelobe gains and even nulls between them.

This paper thus focuses on modeling the cellular-enabled UAV communication system and analyzing its 3D coverage performance for both the uplink and downlink, by taking into account the unique air-ground channel characteristics and  the practical non-uniform GBS antenna pattern.
Specifically, we consider a practical UAV-GBS association strategy where the UAV is associated with the GBS that provides the strongest channel gain with it, which can be implemented by comparing the \textit{reference signal received power} (RSRP) of the downlink signals sent by the GBSs.
To capture the essential feature of air-ground channel, 
we adopt a simplified but practical binary channel state model, comprising only the two states of  LoS or non-LoS (NLoS) with different probabilities of occurrence\cite{3GPPevaluationAssumption}.
As a result, the UAV might be associated with a distant instead of nearby GBS, depending on its channel states with the GBSs as well as angles with their antennas.
Besides GBS association, the coverage performance also depends on the interference with co-channel GBSs, which is a discrete random variable (RV) (as opposed to continuous RV in terrestrial communication) due to the probabilistic LoS/NLoS channel model and the random on/off activities of co-channel GBSs.
As a result, the 3D coverage analysis invokes discrete channel and interference states, and their numbers increase exponentially with the number of involved GBSs within the UAV's signal coverage, which is practically large due to the high probability of LoS.   
To our best knowledge, the coverage performance of the cellular network under such large discrete channel/interference states has not been addressed, which is fundamentally different from that of the terrestrial 2D network with continuous fading channel/interference\cite{AndrewsCellular}, or the 3D air-ground network\cite{ZhangWeiSmallCellJSAC}\cite{Dhillon2017} without considering the probabilistic LoS/NLoS channel.
The main contributions of this paper are summarized as follows.

\begin{itemize}
\item First,
we present a 3D system model for the UAV-GBS uplink/downlink communications, which includes the cellular network model, the 3D patterns of the GBS and UAV antennas, and the 3D air-ground channel. Note that our model is applicable to any given 3D patterns of the GBS/UAV antennas.

\item Second, we propose an analytical framework to characterize the uplink/downlink 3D coverage (average non-outage probability) performance of the UAVs. The new contributions mainly include the consideration of the probabilistic LoS/NLoS channel states, the UAV-GBS association and coupled downlink interference analysis, and the resulted discrete SNR distribution characterization.
To this end, an efficient sorting algorithm is proposed to analyze the UAV-GBS association and uplink SNR distribution, which significantly reduces the complexity from an exponential order with the number of involved GBSs by exhaustive search to a linear order.
Moreover, to resolve the coupling between the UAV-GBS association and downlink interference, the downlink SNR distribution characterization is reduced to deriving the interference distribution conditioned on a given UAV-GBS association.

\item Third, we model the conditional downlink discrete interference given the associated GBS as a new distribution termed as \textit{generalized Poisson multinomial (GPM)}, which take into account the LoS/NLoS channel states and random on/off channel activities of all co-channel GBSs.
However, the size of the sample space of the discrete GPM RV increases exponentially with the number of co-channel GBSs, and as a result its cumulative distribution function (cdf) is prohibitive to compute via the brute-force enumeration-based method. Furthermore, the conditional interference distribution needs to be evaluated over all possible associated GBSs to obtain the downlink SNR distribution and hence the coverage probability.
To reduce such high complexity, we propose a new and efficient method to obtain the conditional interference distribution with high accuracy, named the \textit{lattice approximation (LA)} method, which converts a general non-lattice distributed GPM RV into a lattice distribution with a bounded size of the value space, and then applies the efficient fast Fourier transform (FFT) on its characteristic function to obtain the approximate interference distribution with high accuracy.

\item Finally, extensive numerical results are provided to validate our analysis and reveal insights for system design.
First, it is shown that the GBS antenna pattern has a significant impact on the spatial distribution of non-outage probability, which is useful for UAV path planning/movement control.
Moreover, a large downtilt angle leads to overall smaller GBS antenna gain for the UAV above the GBS height, which affects the uplink and downlink coverage probabilities in different ways.
Second, as the UAV altitude increases, the UAV-GBS link distance increases while so does the LoS probability, both of which affect the link strength and hence the coverage probability, but in opposite ways.
In addition, a high LoS probability at high altitude leads to small variation of the coverage probability versus the SNR threshold.
Third, a high network loading factor with more terrestrial UEs reduces the UAV coverage probability as well as its variation.
Finally, besides the antenna gain, applying directional antenna at the UAV limits the coverage range of its antenna mainlobe and hence the number of covered GBSs, which effectively mitigates the interference at high UAV altitude and thus improves the coverage performance.
\end{itemize}

The rest of this paper is organized as follows.
Section \ref{SectionModel} introduces the 3D system model.
The 3D coverage analysis for the uplink/downlink communication is presented in Section \ref{SectionUplink} and Section \ref{SectionDownlinkMain}, respectively.
Numerical results are provided in Section \ref{SectionSimulation}, followed by conclusions in Section \ref{SectionConclusion}.

\textit{Notations}: $\mathbb{R}$ denotes the set of real numbers; $\mathbb{Z}$ denotes the set of integer numbers; $\mathbb{P}\{\cdot\}$ denotes the probability of an event; $\mathbb{E}\{\cdot\}$ denotes the expectation of an RV; $\|\cdot\|$ denotes the Euclidean norm; $|\cdot|$ denotes the cardinality of a set; $\setminus\cdot$ denotes the set minus operation; $\bigcup$ denotes the set union; $\bigcap$ denotes the set intersection; and $\emptyset$ denotes the empty set.

\section{System Model}\label{SectionModel}

\subsection{Cellular Network Model}\label{SectionCellularModel}
\begin{figure}
        \centering
        \hspace{-10pt}
        \begin{subfigure}[b]{0.56\linewidth}
                \includegraphics[width=1\linewidth,  trim=195 40 125 30,clip]{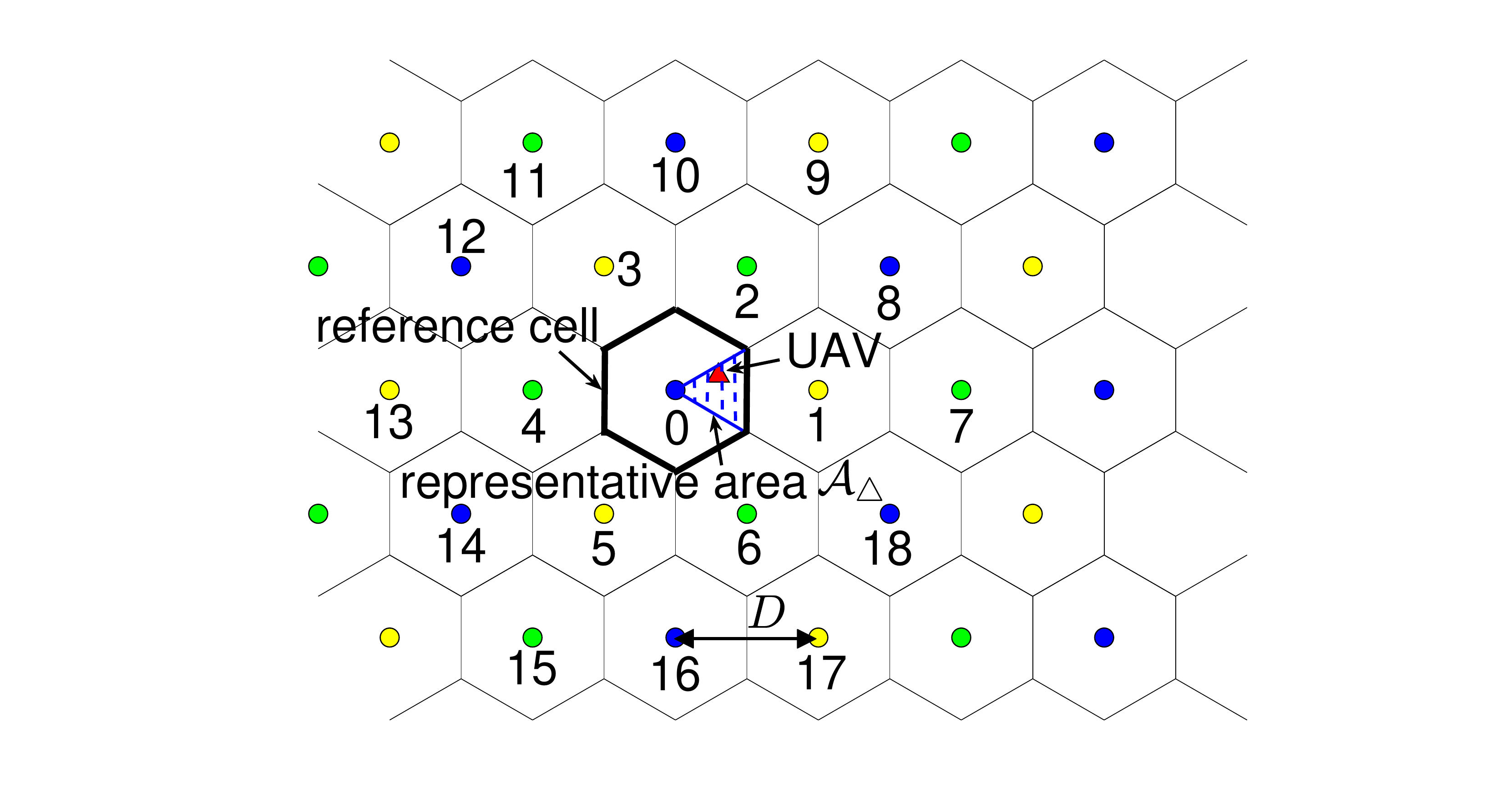}
\caption{\vspace{-2ex}}\label{Hexagon}
        \end{subfigure}%
        \begin{subfigure}[b]{0.53\linewidth}
                \includegraphics[width=1\linewidth,  trim=250 55 125 35,clip]{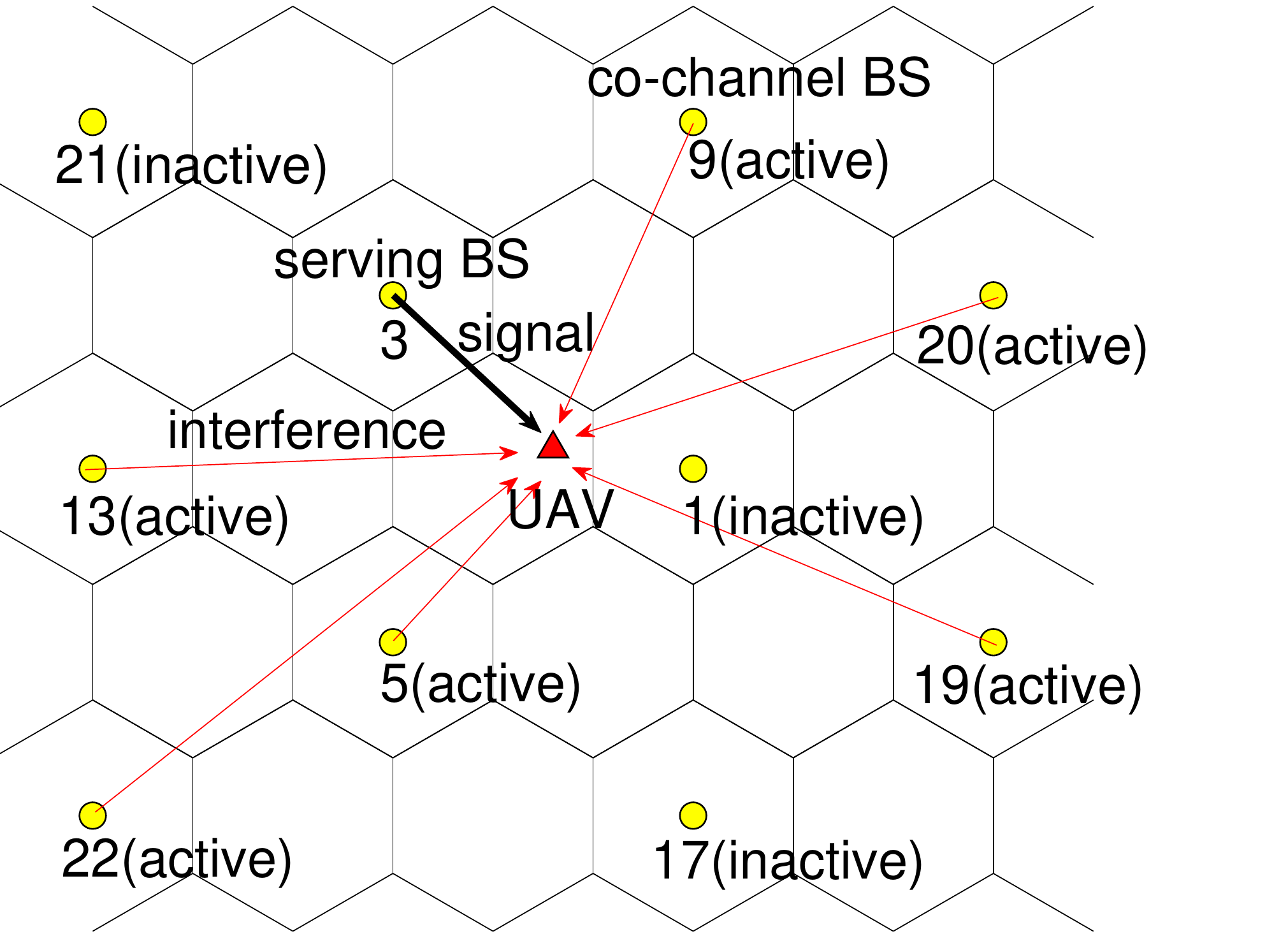}
\caption{\vspace{-2ex}}\label{HexagonDL}
        \end{subfigure}%
        \caption{(a) Hexagon cell layout with frequrency reuse factor $\rho=1/3$. (b) Serving GBS for the UAV and other co-channel GBSs which are active in communicating with their ground UEs and thus generate the downlink interference to the UAV.\vspace{-2ex}}\label{CellLayout}
\end{figure}

Consider a cellular network with the classic hexagonal grid cell layout\footnote{Note that although we use the regular grid as an example, our analysis can be extended to arbitrary cell topology in practice.} where each GBS is at the center of its cell with the inter-cell distance of $D$ meters (m) and a given frequency reuse factor of $\rho=1/\textrm{F}$, where $\textrm{F}\geq 3$ and $\textrm{F}\in \mathbb{Z}$. An example of the network with $\rho=1/3$ is shown in Fig. \ref{CellLayout}(a), where the set of GBSs are denoted as $\mathcal{B}=\{0,1,2,\cdots\}$ and represented by circles of different colors. The whole spectrum is equally divided into $\textrm{F}=3$ orthogonal bands, each of which is reused by the GBSs of the same color, denoted by the co-channel GBS sets $\mathcal{B}_1=\{1,3,5,\cdots\}$ (yellow), and $\mathcal{B}_2=\{2,4,6,\cdots\}$ (green) and $\mathcal{B}_3=\{0,8,10,\cdots\}$ (blue), respectively.
Note that generally $\mathcal{B}_\textrm{f}, \textrm{f}=1,\cdots,\textrm{F}$ are orthogonal and $\bigcup_{\textrm{f}=1}^\textrm{F} \mathcal{B}_\textrm{f}=\mathcal{B}$.
For simplicity, we assume that all GBSs are at an identical height of $H_b$ m, and the 3D coordinate of GBS $i\in\mathcal{B}$ is denoted by $\bold w_i\triangleq(X_i,Y_i,H_b)$, where $X_i,Y_i\in\mathbb{R}$.
Consider a typical UAV user flying at an altitude of $H_u$ m, with $H_u>H_b$, whose 3D coordinate is denoted by $\bold u\triangleq (x_u,y_u,H_u)$, where $x_u,y_u\in\mathbb{R}$. Without loss of generality, we assume that GBS 0 is at the origin and the UAV's horizontal position is randomly located inside the reference cell 0.


\begin{figure}
\centering
   \includegraphics[width=0.8\linewidth]{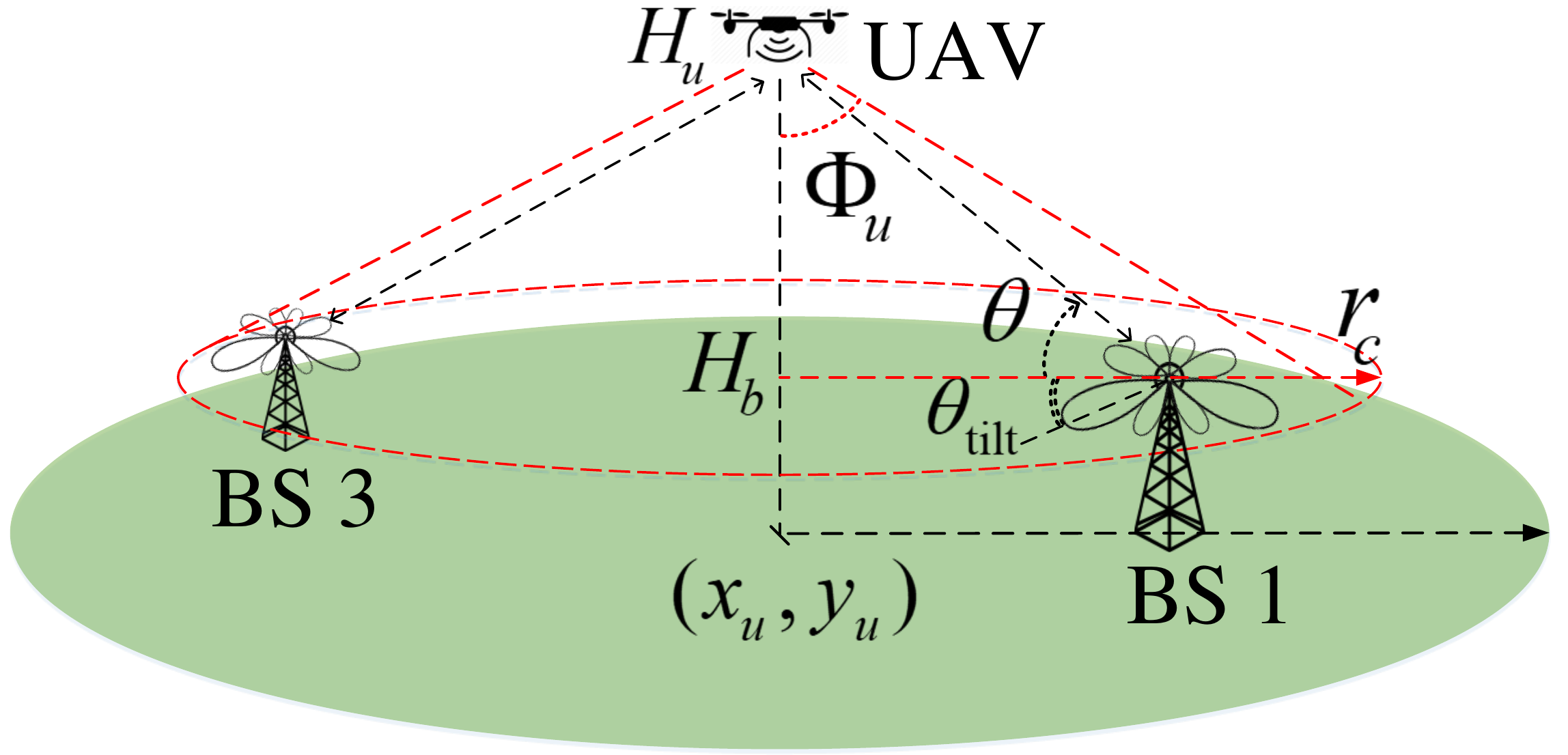}
\caption{Network-connected UAV at high altitude.\vspace{-2ex}}\label{Schematic}
\end{figure}

In the conventional cellular network, each ground UE is usually associated with one of its nearby GBSs for communication based on the average channel gain mainly determined by the distance-dependent pathloss and shadowing. 
However, for the UAV user flying at high altitude, it is possible that the UAV connects to a distant GBS, due to the random realization of LoS/NLoS links, as well as the variation of the GBS antenna radiation pattern in the elevation domain, especially the sidelobes and the nulls in between.
An illustrative example is given in Fig. \ref{Schematic}, where the UAV user is served by the more distant GBS 3 via a strong antenna sidelobe instead of the nearby GBS 1 with a possible null of antenna gain. 
Denote $i_s\in \mathcal{B}$ as the index of the GBS that serves the UAV of interest, and $\textrm{f}_s$ as the index of the set of co-channel GBSs of GBS $i_s$, i.e., $i_s\in \mathcal{B}_{\textrm{f}_s}$. 
For example, suppose that the serving GBS is $i_s=3$, then the set of co-channel GBSs is $\mathcal{B}_{\textrm{f}_s}=\mathcal{B}_1$ as represented by yellow circles in Fig. \ref{CellLayout}(b).

We consider that orthogonal time-frequency resource blocks (RBs) are assigned to the UAV for its uplink and downlink communications, respectively, by its associated serving GBS $i_s$. We assume that the RBs are assigned to users by the GBSs in each co-channel GBS set $\mathcal{B}_\textrm{f}$, independently.
For each co-channel GBS $i\in\mathcal{B}_{\textrm{f}_s}\setminus i_s$,
we assume that the RBs assigned to the UAV by GBS $i_s$ in the uplink/downlink are simultaneously used to serve a ground UE with probability $\omega_{\textrm{ul},i}$ with $0<\omega_{\textrm{ul},i}< 1$, and $\omega_{\textrm{dl},i}$ with $0<\omega_{\textrm{dl},i}< 1$, respectively. The channel active probabilities $\omega_{\textrm{ul},i}$ and $\omega_{\textrm{dl},i}$ reflect the current uplink/downlink \textit{loading factor} of GBS $i$, respectively, which are assumed to be given. 

Specifically, we define a binary variable $\mu_{i}=1$ to indicate that the co-channel GBS $i\in\mathcal{B}_{\textrm{f}_s}\setminus i_s$ is active in communication with a ground UE in the same assigned RB as that of the UAV in the downlink, and otherwise $\mu_{i}=0$. 
Then the downlink $\mu_i$'s for different co-channel GBSs $i\in\mathcal{B}_{\textrm{f}_s}\setminus i_s$ are independent Bernoulli RVs with parameter $\omega_{\textrm{dl},i}$, respectively. Similarly, we can define another binary variable $\nu_i$ to represent the channel activity of co-channel GBS $i\in\mathcal{B}_{\textrm{f}_s}\setminus i_s$ on the uplink RB used by the UAV.


\subsection{Antenna Model}

\begin{figure}
        \centering
        \begin{subfigure}[b]{0.46\linewidth}
                \includegraphics[width=1\linewidth,  trim=260 150 150 110,clip]{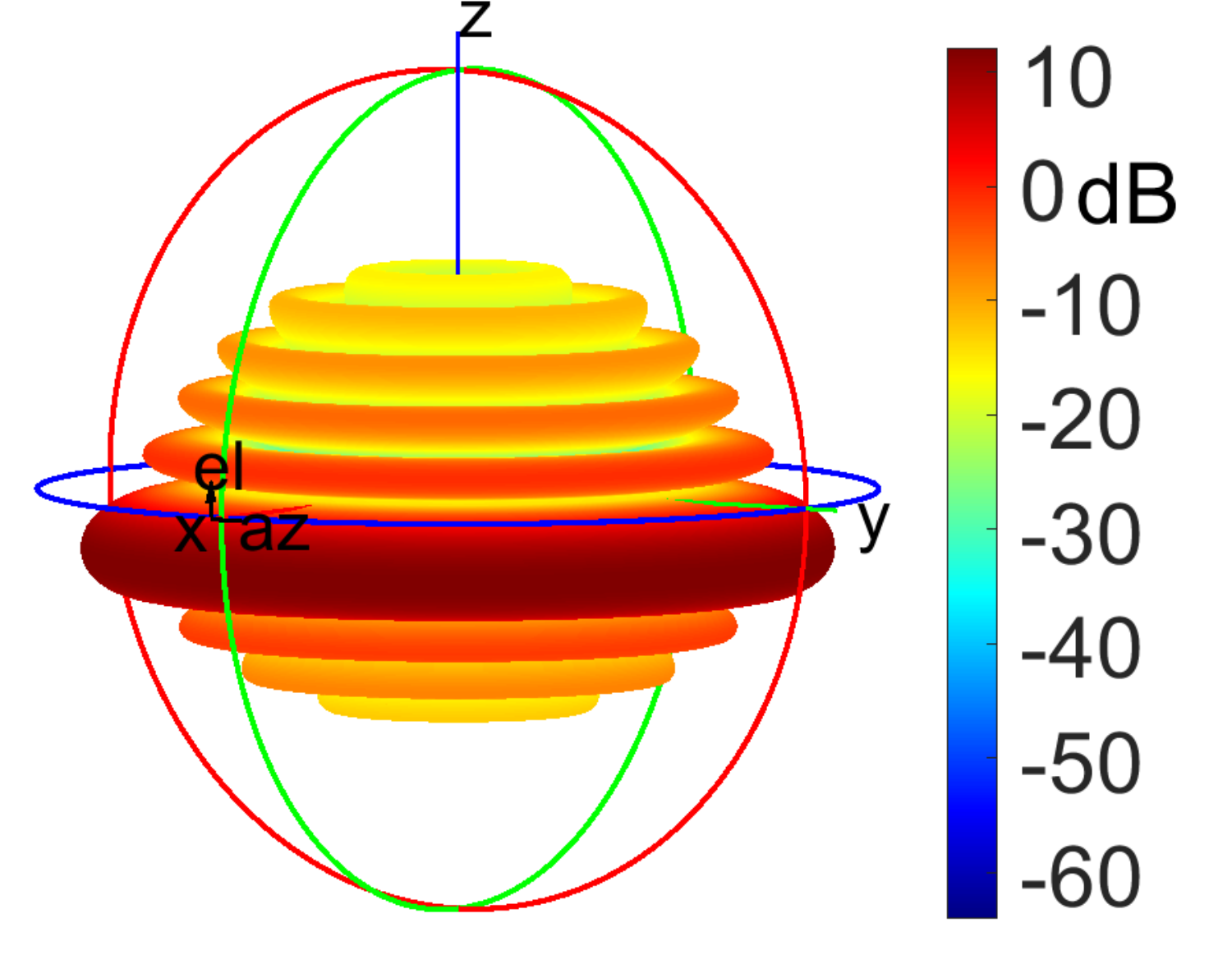}
                \caption{}
                \label{Pattern3D}
        \end{subfigure}%
        \begin{subfigure}[b]{0.27\linewidth}
                \includegraphics[width=1\linewidth,  trim=320 20 300 10,clip]{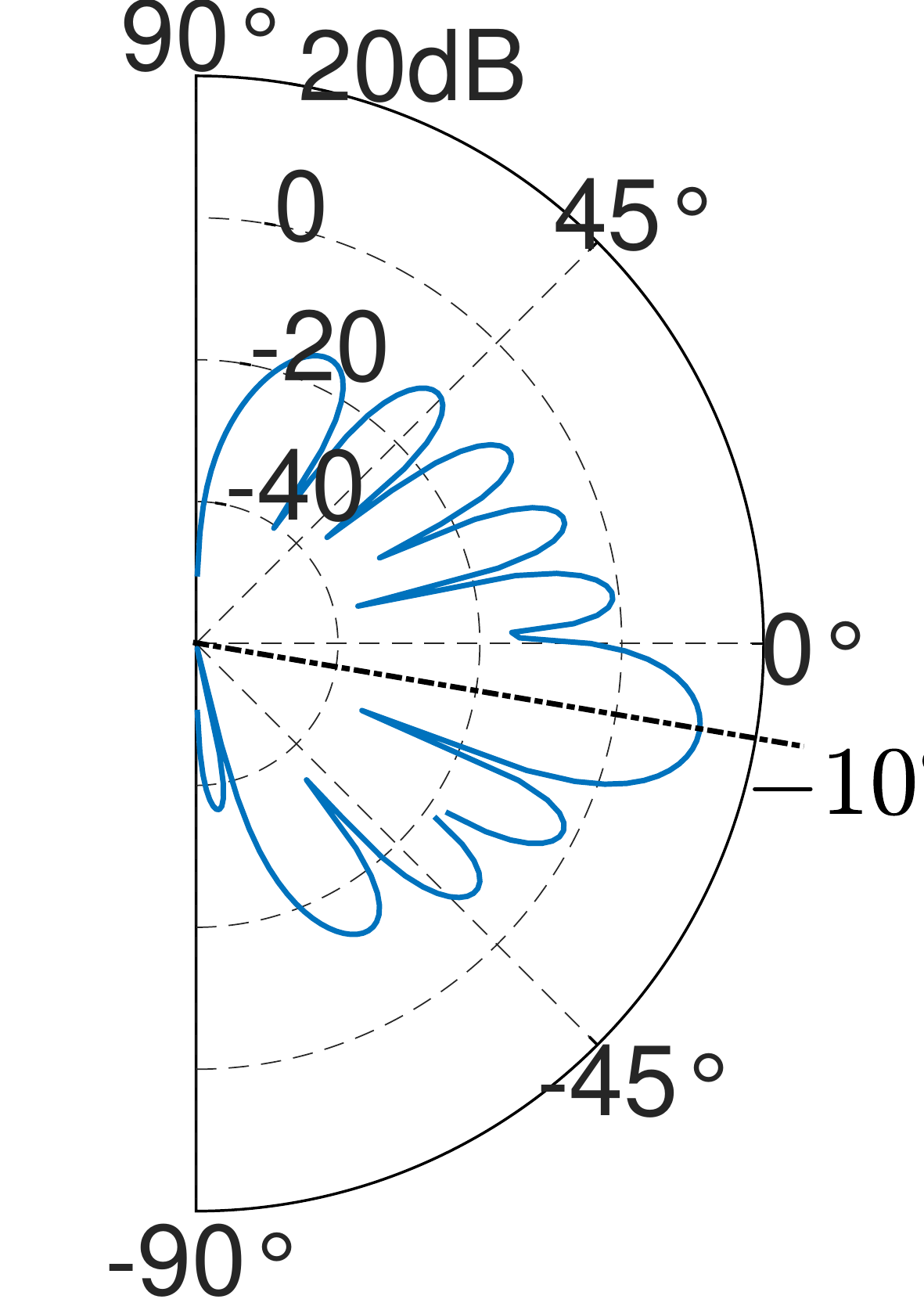}
                \caption{}
                \label{ULAelevation}
        \end{subfigure}%
        \begin{subfigure}[b]{0.27\linewidth}
                \includegraphics[width=1\linewidth,  trim=320 20 300 10,clip]{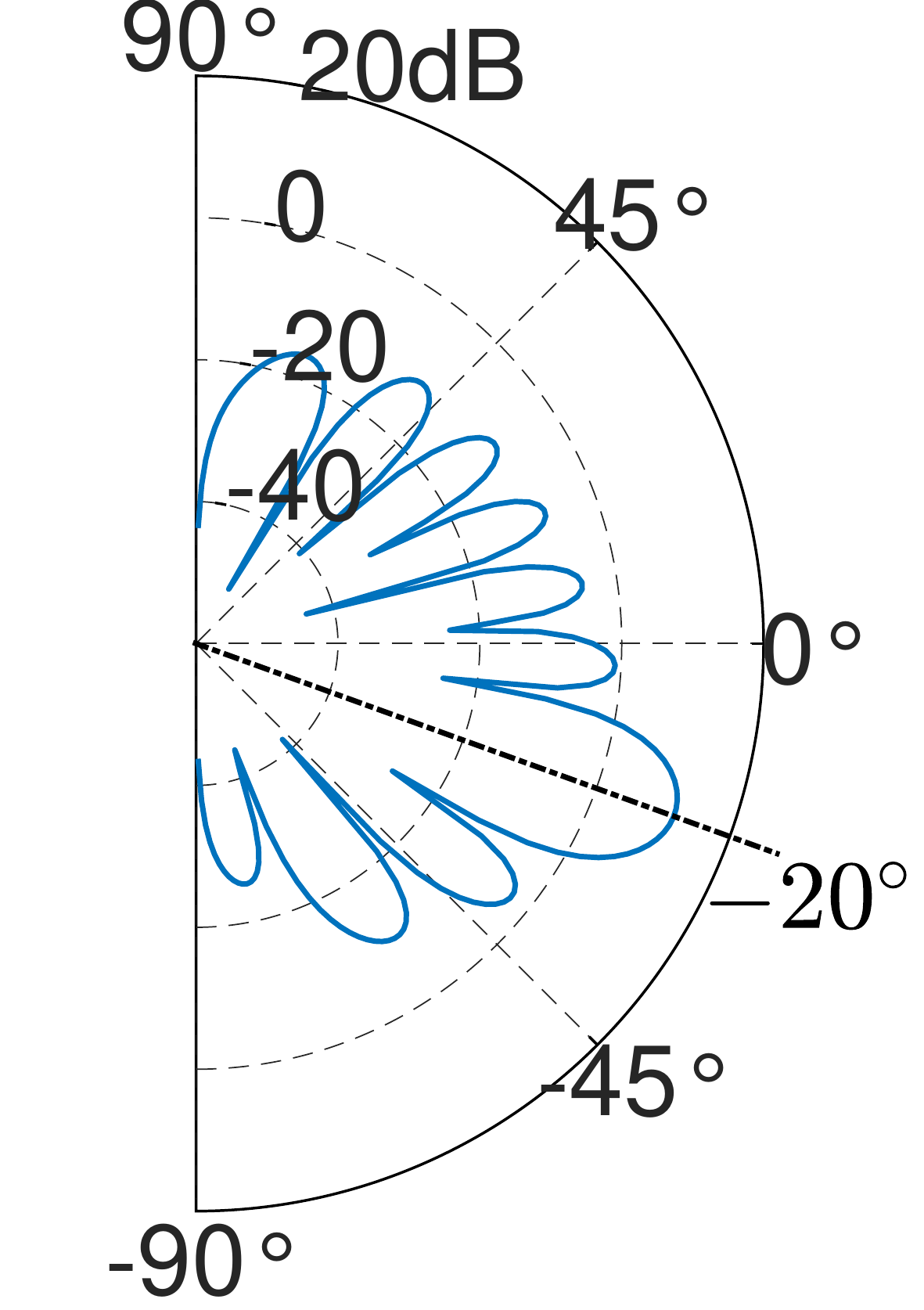}
                \caption{}
                \label{ULAelevation}
        \end{subfigure}%
        \caption{GBS antenna synthesised by the ULA: (a) 3D power gain pattern ($\theta_\textrm{tilt}=-10^\circ$); (b) Elevation pattern ($\theta_\textrm{tilt}=-10^\circ$); (c) Elevation pattern ($\theta_\textrm{tilt}=-20^\circ$).\vspace{-2ex}}\label{BSantennaPlot}
\end{figure}

We consider a practical GBS antenna gain pattern which is omnidirectional in the horizontal plane but vertically directional.\footnote{Our analysis can be readily extended to the case with sectorized antenna pattern of the GBS in the horizontal plane.} Denote $\theta\in(-90^\circ,90^\circ]$ as the elevation angle \textit{upward} from the horizontal plane of the GBS antenna, as shown in Fig. \ref{Schematic}.
Denote the GBS antenna gain at an elevation angle $\theta$ as $G_b(\theta)$. 
For a GBS $i\in\mathcal{B}$, the elevation angle $\theta_i$ as seen by the UAV above the GBS height is given by
\begin{equation}\label{theta_tan}
\theta_i(\bold u)=\arcsin\frac{H_u-H_b}{\|\bold u-\bold w_i\|},\quad \theta_i(\bold u)\in(0^\circ, 90^\circ].
\end{equation}
The specific GBS antenna gain pattern depends on the GBS antenna type and configuration. In existing cellular networks, the GBS antenna is usually tilted downward to support ground UEs, where the antenna boresight direction is electrically or mechanically downtilted with an elevation angle $\theta_{\textrm{tilt}}$ degree $(\theta_{\textrm{tilt}}<0)$, as shown in Fig. \ref{Schematic}.
For the purpose of exposition, we consider in this paper the GBS antenna pattern synthesized by a uniform linear array (ULA)\cite{balanis2016antenna} with $K$ co-polarized dipole antenna elements placed vertically with $d_e$ spacing between elements and electrically steered with downtilt angle $\theta_{\textrm{tilt}}$.
According to \cite{balanis2016antenna}, the power gain pattern of the ULA is given by
\begin{equation}\label{BSantenna}
G_b(\theta)\triangleq G_e(\theta)\big(J(\theta)\big)^2=G_{e,\textrm{max}}\cos^2\theta \bigg(\frac{\sin(\frac{K}{2}\vartheta)}{\sqrt{K}\sin(\frac{1}{2}\vartheta)}\bigg)^2,
\end{equation}
where $\theta\in(-90^\circ,90^\circ]$; $G_e(\theta)\triangleq G_{e,\textrm{max}}\cos^2\theta$ is the power gain pattern of each dipole antenna element with $G_{e,\textrm{max}}$ denoting its maximum value; and $J(\theta)\triangleq\frac{\sin(\frac{K}{2}\vartheta)}{\sqrt{K}\sin(\frac{1}{2}\vartheta)}$ is the normalized array factor of the ULA with $\vartheta\triangleq \frac{2\pi}{\lambda}d_e(\sin\theta-\sin\theta_{\textrm{tilt}})$ in radian (rad) and $\lambda$ being the wavelength. Note that in the downtilt direction, i.e., $\theta=\theta_{\textrm{tilt}}$, we have $\vartheta=0$ and hence
\begin{equation}
J(\theta_{\textrm{tilt}})=\lim\limits_{\vartheta\rightarrow 0}\frac{\sin(\frac{K}{2}\vartheta)}{\sqrt{K}\sin(\frac{1}{2}\vartheta)}=\lim\limits_{\vartheta\rightarrow 0}\frac{\frac{\diff}{\diff \vartheta}\big[\sin(\frac{K}{2}\vartheta)\big]}{\frac{\diff}{\diff \vartheta}\big[\sqrt{K}\sin(\frac{1}{2}\vartheta)\big]}=\sqrt{K},
\end{equation}
which follows from the L'Hospital's rule to evaluate limits.
Therefore, we have
\begin{equation}\label{BSantennaTilt}
G_b(\theta_{\textrm{tilt}})=K G_{e,\textrm{max}}\cos^2\theta_{\textrm{tilt}},
\end{equation}%
which approximately achieves the maximum antenna gain in \eqref{BSantenna} for small downtilt angle in practice, e.g., $|\theta_{\textrm{tilt}}|\leq 20^\circ$.
For illustration, the power gain pattern of the ULA in the 3D space for $\theta_{\textrm{tilt}}=-10^\circ$ is plotted in Fig. \ref{BSantennaPlot} (a), where the half-wave dipole element is used with $G_{e,\textrm{max}}=1.64$ and the following parameters: $K=10$, $d_e=0.5\lambda$ and carrier frequency $f_c=2$ GHz. The corresponding 2D patterns in the elevation domain for $\theta_{\textrm{tilt}}=-10^\circ$ and $\theta_{\textrm{tilt}}=-20^\circ$ are plotted in Fig. \ref{BSantennaPlot} (b) and (c), respectively.
Similar examples of GBS antenna pattern synthesized by antenna arrays can be found in the latest 3GPP technical report\cite{3GPP3Dmodel}. Note that our proposed analytical framework in this paper is general and can be applied to any given GBS antenna pattern. 

On the other hand, the power gain pattern of the UAV antenna also plays an important role in the link performance. In particular, the use of directional antenna with different boresight directions at different UAV altitudes can effectively confine the interfered/interfering co-channel GBSs in the uplink/downlink within a limited region and hence alleviate the severe air-ground interference issue due to strong LoS channels.
Note that depending on the hardware configuration and flying altitude of the UAV, the boresight direction of the UAV antenna in general could be adjustable within a certain range. For simplicity, in this paper, we consider a given boresight direction of the UAV antenna at each altitude.
For the purpose of exposition, we assume that the UAV is equipped with a directional antenna whose boresight direction is pointing downward to the ground, and the azimuth and elevation half-power beamwidths are both $2\Phi_u$ degrees (deg) with $\Phi_u\in(0^\circ,90^\circ)$, as shown in Fig. \ref{Schematic}. Furthermore, the corresponding antenna power gain in direction $(\phi_a,\phi_e)$ can be practically approximated as
\begin{align}\label{UAVantenna}
G_u(\phi_a,\phi_e)=
\begin{cases}
G_0/\Phi_u^2, & \textrm{$\phi_a,\phi_e\in[-\Phi_u,\Phi_u]$;}\\
g_0\approx 0, & \ \textrm{otherwise,}
\end{cases}
\end{align}%
where $G_0=\frac{30000}{2^2}=7500$; $\phi_a$ and $\phi_e$ denote the azimuth and elevation angles from the antenna boresight direction, respectively \cite{balanis2016antenna}. Note that in practice, $g_0$ satisfies $0<g_0 \ll G_0/\Phi_u^2$, and for simplicity we assume $g_0= 0$ in this paper.
Further note that, in the special case with $\Phi_u=90^\circ$, we have $G_u(\phi_a,\phi_e)\approx 1, \forall \phi_a, \phi_e$, which reduces to the case with an isotropic antenna.
The antenna gain of the UAV as seen by a GBS $i\in\mathcal{B}$ is thus given by
\begin{align}\label{UAVantennaBS0}
G_{u,i}(\bold u)=
\begin{cases}
G_0/\Phi_u^2, & \textrm{$d_i(\bold u)\leq r_c$;}\\
0, & \ \textrm{otherwise,}
\end{cases}
\end{align}%
where $d_i(\bold u)\triangleq \sqrt{(x_u-X_i)^2+(y_u-Y_i)^2}$ is the horizontal distance from the UAV to GBS $i$, and $r_c\triangleq(H_u-H_b)\tan\Phi_u$ is the radius of the \textit{coverage area} of the UAV antenna main-lobe projected on the horizontal plane at the GBS height, as shown in Fig. \ref{Schematic}. 
As a result, the UAV can only communicate/interfere with the GBSs within its antenna coverage area.

\subsection{Channel Model}
Due to the high altitude of the UAV, LoS channel exists with a high probability for practical UAV-GBS links\cite{3GPPevaluationAssumption}, which has been experimentally verified (see e.g., the recent field measurement report by Qualcomm\cite{QualCommDroneReport}).
According to \cite{AirToGroundThreeGroup}, the received signals in UAV-ground communications mainly constitute three components, namely LoS signal, strong reflected NLoS signals and multiple reflected signals which cause multi-path fading, each with a certain probability of occurrence.
Typically, the probability of receiving LoS and strong NLoS signals are significantly higher than that of multi-path fading \cite{AirToGroundVTC}. 
Therefore, for simplicity as well as capturing the main characteristic of UAV-ground channels, we ignore the multi-path fading in this paper and consider only the dominant LoS and NLoS components. 

In practice, the LoS/NLoS probability and associated pathloss depend on the density and height of buildings in a given environment, as well as the relative position between the UAV and GBS. Therefore, in a given environment, the channel power gain between GBS $i\in\mathcal{B}$ and the UAV at position $\bold u$ can be expressed as
\begin{align}\label{probLOS}
h_i(\bold u)=
\begin{cases}
h_{\textrm{L},i}(\bold u), & \quad \textrm{with probability $p_{\textrm{L},i}(\bold u)$;}\\
h_{\textrm{NL},i}(\bold u), & \quad \textrm{with probability $p_{\textrm{NL},i}(\bold u)$,}
\end{cases}
\end{align}%
where $h_{\textrm{L},i}$ and $p_{\textrm{L},i}$ denote the channel power gain and occurrence probability of the LoS channel, respectively, while $h_{\textrm{NL},i}$ and $p_{\textrm{NL},i}$ denote the counterparts for the NLoS channel, respectively, and $p_{\textrm{NL},i}=1-p_{\textrm{L},i}$.
For the specific forms of the involved functions in \eqref{probLOS}, interested readers may refer to the simplified formula in \cite{LAPlosProbability} or the empirical formula in \cite{3GPPevaluationAssumption}. 
Furthermore, we define a binary variable $\delta_i$ to represent the event of LoS/NLoS channel occurrence, i.e.,
\begin{align}\label{BinaryChannel}
\delta_{i}\triangleq
\begin{cases}
1, & \textrm{LoS channel between UAV and GBS $i$;}\\
0, & \textrm{NLoS channel between UAV and GBS $i$.}
\end{cases}
\end{align}%
Then the $\delta_i$'s for different GBSs $i\in\mathcal{B}$ are independent Bernoulli RVs with parameter $p_{\textrm{L},i}$, respectively.
Then the channel gain in \eqref{probLOS} can be rewritten as 
\begin{equation}\label{probLOSoneline}
h_i(\bold u)=\delta_i h_{\textrm{L},i}(\bold u)+ (1-\delta_i) h_{\textrm{NL},i}(\bold u).
\end{equation}

For simplicity in this paper, we consider two states (LoS/NLoS) for the channel gain $h_i(\bold u)$ in \eqref{probLOSoneline}, which has not modeled the shadowing gain variation with UAV locations in the NLoS channel case.
The authors in \cite{UAVradioMapSegmented} proposed a segmented regression approach to further extend the channel model to constitute multiple propagation groups besides LoS/NLoS (e.g., obstructed LoS), which is shown to be able to reconstruct the segmented structure of the UAV-ground propagation conditions observed in practice and by ray tracing simulations.
It is worth pointing out that our proposed analytical framework is extendable to other channel models, with any finite number of channel states. For example, for the case of NLoS, we can add a shadowing gain factor $\xi$ (e.g., log-normal random variable) to the NLoS power gain $h_{\textrm{NL},i}(\bold u)$, and then quantize this shadowing gain distribution approximately into a discrete random variable with finite number of states. As such, our proposed analytical method still applies, as we consider generalized multinomial distributions that can be used to model any finite number of channel states.

\subsection{UAV-GBS Association}

Denote $C_i(\bold u)$ as the overall power gain by taking into account both the channel power gain and the antenna power gains of the UAV and GBS $i\in\mathcal{B}$, which is given by
\begin{align}\label{CombinedGainGeneral}
&C_i(\bold u)\triangleq G_{u,i}(\bold u) h_i(\bold u) G_b\big(\theta_i(\bold u)\big)\notag\\
&=G_{u,i}(\bold u) G_b\big(\theta_i(\bold u)\big)\big(\delta_i h_{\textrm{L},i}(\bold u)+ (1-\delta_i) h_{\textrm{NL},i}(\bold u)\big).
\end{align}
In this paper, we consider a practical user association rule where the UAV is associated with the GBS that provides the strongest power gain with it,\footnote{Here we assume that all GBSs have available RBs to support the uplink/downlink communications of a new UAV user.} which can be implemented by comparing the RSRP of the downlink beacon signals sent by the GBSs in $\mathcal{B}$. Assume that the beacon signals are sent using the same transmit power. Then the serving GBS $i_s$ is selected as the one with the largest power gain, i.e.,
\begin{equation}
i_s\triangleq \arg\max\limits_{i\in\mathcal{B}} C_i(\bold u).
\end{equation}
Accordingly, the handover of the UAV between different GBSs can follow the typical procedure in cellular networks. Specifically, the UAV continually monitors the RSRP of the GBSs it can hear, including the one it is currently associated with, and feeds this information back. When the RSRP from the serving GBS starts to fall below a certain level, the cellular network looks at the RSRP from other GBSs reported by the UAV, and makes the decision whether to handover or not and to which GBS.

Note that due to the high altitude, UAV can potentially have strong LoS links with a large number of co-channel GBSs. As a result, it is difficult to schedule/reserve a dedicated RB for the UAV's exclusive use, or perform inter-cell coordination/interference cancellation for the UAV user as they require centralized control over many GBSs within the UAV's wide communication range. This is especially the case with a high network-loading factor where the number of active ground UEs in each cell is already large, and thus the reuse of RBs is inevitable. Therefore, this paper aims to provide a baseline performance analysis for the UAV's uplink/downlink communications without the need of centralized inter-cell scheduling/coordination.

\subsection{Uplink Communication}\label{SectionSystemUplink}

Besides the direct communication link with the associated GBS, the interference links also affect the 3D coverage performance of the UAV.
Specifically, there are two new types of interference in a cellular network supporting both ground UEs and UAV users, which are respectively the interference between UAV users and ground UEs, and that among different UAVs even when they are not densely distributed.
In this paper, we assume that different UAVs are assigned with orthogonal channels and thus there is no inter-UAV interference.
We further assume that the traditional interference issue among ground UEs in co-channel cells is effectively resolved by existing techniques such as cell planning, frequency reuse, dynamic RB allocation, power control, beamforming, etc.
Therefore, we focus on studying the UAV's uplink/downlink performance, which is the new contribution of this work.



For uplink communication, we assume that the UAV transmits with power $P_u$ in Watt (W), which is capped by the maximum transmit power $P_{\textrm{max}}$.
The received SNR\footnote{We consider SNR here without any interference from ground UEs, since such interference belongs to the conventional terrestrial interference in the cellular uplink, which we have assumed to be negligible and treated as background noise.} at the receiver of the serving GBS $i_s$ is thus given by
\begin{equation}\label{uplinkSNR}
\gamma_{\textrm{ul}}(\bold u)\triangleq\frac{P_u C_{i_s}(\bold u)}{\sigma^2}=\beta_0\max\limits_{i\in\mathcal{B}} C_i(\bold u),
\end{equation}
where $\beta_0\triangleq P_u/\sigma^2$ and the receiver noise is assumed to be additive white Gaussian noise (AWGN) with zero mean and power $\sigma^2$.

Note that the uplink SNR $\gamma_{\textrm{ul}}(\bold u)$ at each given UAV location $\bold u$ is an RV due to the probabilistic LoS channel occurrence $\delta_i, i\in\mathcal{B}$.
Moreover, $\gamma_{\textrm{ul}}(\bold u)$ also depends on the uplink transmit power $P_u$. 
We consider that the UAV has an uplink SNR requirement of $\gamma_{\textrm{ul}}(\bold u)\geq \eta_\textrm{ul}$ in order to be in \textit{non-outage}, where $\eta_\textrm{ul}>0$ is a pre-defined SNR threshold.
The uplink \textit{outage probability} of the UAV at location $\bold u$ is thus defined as
\begin{equation}\label{PoutUplink}
p_\textrm{out,ul}(\bold u)\triangleq\mathbb{P}\{\gamma_{\textrm{ul}}(\bold u)<\eta_\textrm{ul}\}=F_{\gamma_{\textrm{ul}}(\bold u)}(\eta_\textrm{ul}),
\end{equation}
where $F_{\gamma_{\textrm{ul}}(\bold u)}(\cdot)$ denotes the cdf\footnote{The cdf of an RV $X$ is defined as $F_{X}(x)\triangleq \mathbb{P}\{X< x\}$.} of $\gamma_{\textrm{ul}}(\bold u)$.

Assume that the UAV performs power control with the objective to satisfy the SNR requirement, while at the same time reducing the interference to co-channel GBSs, subject to the maximum transmit power $P_\textrm{max}$.
Specifically,
in the case where the channel gain $C_{i_s}$ of the serving GBS $i_s$ is sufficiently large such that $\frac{P_\textrm{max} C_{i_s}(\bold u)}{\sigma^2}\geq \eta_\textrm{ul}$, then the SNR requirement can be satisfied and the UAV can reduce its transmit power to $P_u=\eta_\textrm{ul}\sigma^2/C_{i_s}(\bold u)$.
On the other hand, in the case with $\frac{P_\textrm{max} C_{i_s}(\bold u)}{\sigma^2}< \eta_\textrm{ul}$, the SNR requirement cannot be satisfied even with the maximum transmit power $P_\textrm{max}$; as a result, the UAV is said to be in the uplink ``coverage hole". In this case, the UAV can keep silent to avoid the interference to other co-channel GBSs. 
In the rest of this paper, without loss of generality, we consider $P_u=P_\textrm{max}$ in order to characterize the (maximum) non-outage probability in the uplink.

In practice, if the UAV is allowed to adjust its position $\bold u$ within a certain region, then the UAV can move to other position $\bold u'$ with potentially better SNR to get out of the coverage hole.
For this purpose, the spatial distribution of the outage probability within the considered region is useful to guide the direction of the UAV movement in practice.
In Section \ref{SectionSimulation}, we will investigate the spatial pattern of outage probability distribution for such applications.

Next, we define the uplink coverage probability of the UAV in the 3D space $\mathcal{S}\subset \mathbb{R}^3$ as the spatial average of non-outage probability over the space, i.e.,
\begin{equation}\label{PcUplink}
p_\textrm{c,ul}(\mathcal{S})\triangleq\mathbb{E}_{\bold u\in\mathcal{S}}\big\{\mathbb{P}\{\gamma_{\textrm{ul}}(\bold u)\geq\eta_{\textrm{ul}}\}\big\}=1-\mathbb{E}_{\bold u\in\mathcal{S}}\big\{p_\textrm{out,ul}(\bold u)\big\},
\end{equation}
where the UAV position $\bold u$ is assumed to be uniformly distributed in the considered 3D space $\mathcal{S}$.
More specifically, we consider two subspaces. First, define the 2D subspace at a given UAV altitude $H_u$ within the 2D horizontal area $\mathcal{A}\subset \mathbb{R}^2$ as 
\begin{equation}
\mathcal{S}(\mathcal{A},H_u)\triangleq \{(x,y)\in \mathcal{A}, z=H_u\}.
\end{equation}
Second, define the 3D subspace between an altitude range $H_u\in [H_\textrm{min},H_\textrm{max}]$ as\footnote{In practice, there are usually limits on the maximum and minimum UAV altitude due to the air traffic control.}  
\begin{equation}
\mathcal{S}(\mathcal{A},[H_\textrm{min},H_\textrm{max}])\triangleq \{(x,y)\in \mathcal{A}, z\in [H_\textrm{min},H_\textrm{max}]\}.
\end{equation}
Accordingly, we can investigate the coverage probability at a certain UAV altitude or altitude range in the cellular network.

In the considered hexagonal cell layout, thanks to symmetry, the coverage probability of the UAV can be obtained by applying \eqref{PcUplink} to the uniformly distributed UAV horizontal positions within the area of the reference cell 0, denoted as $\mathcal{A}_0$.
Note that by symmetry, we can further divide each reference cell into six centrally-symmetric triangular parts, and then we only need to average over one such part $\mathcal{A}_{\triangle}$ as shown in Fig. \ref{CellLayout}(a) to compute the coverage probability.
Therefore, the overall uplink coverage probability of the UAV at an altitude $H_u$ is given by
\begin{equation}\label{PcUL2d}
p_\textrm{c,ul}\big(\mathcal{S}(\mathcal{A}_\triangle,H_u)\big)=1-\mathbb{E}_{\bold u\in \mathcal{S}(\mathcal{A}_\triangle,H_u)}\big\{p_\textrm{out,ul}(\bold u)\big\}.
\end{equation}
The overall uplink coverage probability of the UAV between an altitude range $[H_\textrm{min},H_\textrm{max}]$ is then given by
\begin{equation}\label{PcUL3d}
p_\textrm{c,ul}\big(\mathcal{S}(\mathcal{A}_\triangle,[H_\textrm{min},H_\textrm{max}])\big)=\int_{H_\textrm{min}}^{H_\textrm{max}} p_\textrm{c,ul}\big(\mathcal{S}(\mathcal{A}_\triangle,H_u)\big) \diff H_u.
\end{equation}

\subsection{Downlink Communication}\label{SectionDownlink}
For downlink communication, the UAV receives interference from potentially a large number of co-channel GBSs (see Fig. \ref{CellLayout}(b)), which cannot be ignored due to the strong LoS channels.
Assume that each GBS transmits with the same power $P_{b}$ (W) to the UAV/ground UE in the donwlink.
Then the desired signal power received by the UAV from the serving GBS $i_s$ is given by $P_{b}C_{i_s}(\bold u)$; while the interference power from each active co-channel GBS $i\in\mathcal{B}_{\textrm{f}_s}\setminus i_s$ is given by $P_{b}I_i(\bold u)$, where $I_i(\bold u)\triangleq\mu_i  C_{i}(\bold u)$.
The aggregate interference power (normalized by $P_b$) received by the UAV is thus given by
\begin{equation}\label{interferenceTotal}
I(\bold u)\triangleq\sum\limits_{i\in\mathcal{B}_{\textrm{f}_s}\setminus i_s} I_i(\bold u)= \sum\limits_{i\in\mathcal{B}_{\textrm{f}_s}\setminus i_s} \mu_i  C_{i}(\bold u).
\end{equation}
As a result, the SNR received at the UAV is given by
\begin{equation}\label{DLsinr}
\gamma_{\textrm{dl}}(\bold u)\triangleq \frac{P_{b}C_{i_s}(\bold u)}{P_{b} I(\bold u)+\sigma^2}=\frac{\max\limits_{i\in\mathcal{B}} C_i(\bold u)}{\alpha_0+\sum\limits_{i\in\mathcal{B}_{\textrm{f}_s}\setminus i_s} \mu_i  C_{i}(\bold u)},
\end{equation}
where $\alpha_0\triangleq\sigma^2/P_b$.
The downlink SNR $\gamma_{\textrm{dl}}(\bold u)$ at UAV location $\bold u$ is an RV due to the random channel activity $\mu_i$ of co-channel GBSs and the probabilistic LoS channel occurrence $\delta_i$ in $C_i, i\in\mathcal{B}$.

We assume that the UAV has a downlink SNR requirement of $\gamma_{\textrm{dl}}(\bold u)\geq \eta_\textrm{dl}$ in non-outage, where $\eta_\textrm{dl}>0$ is a pre-defined SNR threshold.
The downlink outage probability of the UAV at location $\bold u$ can then be defined as
\begin{equation}\label{PoutDownlink}
p_\textrm{out,dl}(\bold u)\triangleq\mathbb{P}\{\gamma_{\textrm{dl}}(\bold u)<\eta_\textrm{dl}\}=F_{\gamma_{\textrm{dl}}(\bold u)}(\eta_\textrm{dl}),
\end{equation}
where $F_{\gamma_{\textrm{dl}}(\bold u)}(\cdot)$ denotes the cdf of $\gamma_{\textrm{dl}}(\bold u)$.
Based on \eqref{PoutDownlink}, we can similarly define $p_\textrm{c,dl}(\mathcal{S})$, $p_\textrm{c,dl}\big(\mathcal{S}(\mathcal{A}_\triangle,H_u)\big)$ and $p_\textrm{c,dl}\big(\mathcal{S}(\mathcal{A}_\triangle,[H_\textrm{min},H_\textrm{max}])\big)$ for the downlink similarly as their uplink counterparts.

\section{UAV-GBS Association and Uplink Outage Analysis}\label{SectionUplink}

In this section, we analyze the UAV's uplink outage probability in \eqref{PoutUplink} at any given location $\bold u$, which is essential to the characterization of the 3D coverage probability in \eqref{PcUplink}. For uplink communication, the SNR $\gamma_{\textrm{ul}}(\bold u)$ is determined by the UAV-GBS association based on the strongest power gain, which is, however, random due to the probabilistic LoS channel occurrence $\delta_i, i\in\mathcal{B}$.

For a given UAV location $\bold u$, suppose that the antenna power gains of the UAV and GBS $i\in\mathcal{B}$, the channel power gains $h_{\textrm{L},i}(\bold u)$ and $h_{\textrm{NL},i}(\bold u)$ of LoS/NLoS channels, and the LoS probability $p_{\textrm{L},i}(\bold u)$ are all given. 
For notation simplicity, we drop $(\bold u)$ in the following analysis. 
Then the uplink SNR in \eqref{uplinkSNR} can be rewritten as
\begin{equation}
\gamma_{\textrm{ul}}=\beta_0\max\limits_{i\in\mathcal{B}} C_i=\beta_0\max\limits_{i\in\mathcal{B}} G_{u,i} G_b\big(\theta_i\big)\big(\delta_i h_{\textrm{L},i}+ (1-\delta_i) h_{\textrm{NL},i}\big),
\end{equation}
where each term $C_i$ can take two possible values of $C_{\textrm{L},i}\triangleq G_{u,i} G_b\big(\theta_i\big)h_{\textrm{L},i}$ or $C_{\textrm{NL},i}\triangleq G_{u,i} G_b\big(\theta_i\big)h_{\textrm{NL},i}$, depending on the realization of the Bernoulli RV $\delta_i$ which takes the value of 1 with probability $p_{\textrm{L},i}$ or 0 with probability $1-p_{\textrm{L},i}$. In general, we have $C_{\textrm{L},i}>C_{\textrm{NL},i}, \forall i\in\mathcal{B}$.

In order to obtain the probability mass function (pmf)\footnote{The pmf of a discrete RV $X$ is defined as $f_{X}(x)\triangleq \mathbb{P}\{X= x\}$.} of the uplink SNR $\gamma_{\textrm{ul}}$, a direct method is to enumerate $2^{|\mathcal{B}|}$ possible combinations of $C_i, i\in\mathcal{B}$, then find for each realization the maximum $C_i$ and its probability of occurrence.
This enumeration-based method has an exponential complexity in terms of $|\mathcal{B}|$, which can be practically large (e.g., when the UAV altitude $H_u$ is high and the UAV antenna beamwidth $\Phi_u$ is large, the UAV can potentially establish strong LoS links with a large number of GBSs\footnote{Note that due to the probabilistic LoS/NLoS channel realization and the non-uniform GBS antenna pattern in the elevation domain, the serving GBS is not necessarily nearby the UAV, but instead can be quite far apart.}). 
In order to reduce such complexity, we propose an efficient algorithm to obtain the pmf of $\gamma_{\textrm{ul}}$ in Algorithm \ref{AlgUplink}. 
The key idea is to sort the LoS channel gains $\{C_{\textrm{L},i}\}_{i\in\mathcal{B}}$ in descending order, then obtain the maximum $C_i$ and its probability one by one based on the sorted order. 
Specifically, denote the ordered index of the GBSs as $i_m, m=1,\cdots,|\mathcal{B}|$. First, we have $\mathbb{P}\{\gamma_{\textrm{ul}}= \beta_0 C_{\textrm{L},i_1}\}=p_{\textrm{L},i_1}$, since $\max\limits_{i\in\mathcal{B}} C_i=C_{\textrm{L},i_1}$ when the channel realization between the UAV and GBS $i_1$ is LoS.
Similarly, we have $\mathbb{P}\{\gamma_{\textrm{ul}}= \beta_0 C_{\textrm{L},i_m}\}=p_{\textrm{L},i_m}\prod_{j=1}^{m-1}(1-p_{\textrm{L},i_j})$, since $\max\limits_{i\in\mathcal{B}} C_i=C_{\textrm{L},i_m}$ when the channel realization between the UAV and GBS $i_m$ is LoS while those between the UAV and GBSs $i_1,\cdots,i_{m-1}$ are NLoS.
As a result, the worse-case complexity of Algorithm \ref{AlgUplink} is only linear in the maximum number of iterations $|\mathcal{B}|$.

To further reduce the number of iterations in Algorithm \ref{AlgUplink}, two early stopping criteria can be applied.
First, let $C_{\textrm{NL}}^{\textrm{max}}\triangleq \max_{i\in\mathcal{B}} C_{\textrm{NL},i}$ be the maximum NLoS channel gain among the GBSs in $\mathcal{B}$. Then the algorithm can stop early if the current LoS channel gain $C_{\textrm{L},i_m}$ is smaller than $C_{\textrm{NL}}^{\textrm{max}}$, since the maximum $C_i$ cannot be smaller than $C_{\textrm{NL}}^{\textrm{max}}$.
Second, in the $m$-th iteration, if the probability term $\prod_{j=1}^{m-1}(1-p_{\textrm{L},i_j})$ is lower than a prescribed small threshold value $\epsilon> 0$, then we can also neglect the rest of iterations and stop the algorithm early.
Finally, after the pmf of uplink SNR $\gamma_{\textrm{ul}}(\bold u)$ at UAV location $\bold u$ is derived, we can then obtain its cdf and hence the uplink outage probability in \eqref{PoutUplink}.

\begin{algorithm}[H]\caption{Computing the pmf of uplink SNR $\gamma_{\textrm{ul}}$}\label{AlgUplink}
\begin{small}
\begin{algorithmic}[1]
\STATE Sort $\{C_{\textrm{L},i}\}_{i\in\mathcal{B}}$ in descending order, and denote the ordered index of the GBSs as $i_m, m=1,\cdots,|\mathcal{B}|$.
\STATE Let $C_{\textrm{NL}}^{\textrm{max}}\triangleq \max_{i\in\mathcal{B}} C_{\textrm{NL},i}$.
\STATE Set $\mathbb{P}\{\gamma_{\textrm{ul}}= \beta_0 C_{\textrm{L},i_1}\}=p_{\textrm{L},i_1}$.
\FOR{$m=2,\cdots,|\mathcal{B}|$}
\IF{$C_{\textrm{L},i_m}<C_{\textrm{NL}}^{\textrm{max}}$}
\STATE Set $\mathbb{P}\{\gamma_{\textrm{ul}}= \beta_0 C_{\textrm{NL}}^{\textrm{max}}\}=\prod_{j=1}^{m-1}(1-p_{\textrm{L},i_j})$;
\STATE go to END.
\ENDIF
\STATE $\mathbb{P}\{\gamma_{\textrm{ul}}= \beta_0 C_{\textrm{L},i_m}\}=p_{\textrm{L},i_m}\prod_{j=1}^{m-1}(1-p_{\textrm{L},i_j})$.
\ENDFOR
\STATE $\mathbb{P}\{\gamma_{\textrm{ul}}= \beta_0 C_{\textrm{NL}}^{\textrm{max}}\}=\prod_{j=1}^{|\mathcal{B}|}(1-p_{\textrm{L},i_j})$.
\STATE END: Set for all other values of $\gamma_{\textrm{ul}}$ probability 0.
\end{algorithmic}
\end{small}
\end{algorithm}


\section{Downlink Outage Analysis}\label{SectionDownlinkMain}
In this section, we analyze the UAV's downlink outage probability in \eqref{PoutDownlink} at any given location $\bold u$, which is more involved as compared to that in the uplink derived in the previous section.
For downlink communication, the SNR $\gamma_{\textrm{dl}}(\bold u)$ depends on not only the UAV-GBS association, but also the aggregate interference $I(\bold u)$ from all co-channel GBSs $i\in\mathcal{B}_{\textrm{f}_s}\setminus i_s$, which is a discrete RV due to the probabilistic LoS channel occurrence $\delta_i$ and the random channel activity $\mu_i$.
First, we derive the general cdf expression of $\gamma_{\textrm{dl}}(\bold u)$, by resolving the coupling between the UAV-GBS association and the aggregate interference based on conditional probabilities. Then conditioned on the associated GBS, we further investigate the cdf of aggregate interference with practically a large number of co-channel GBSs in our considered system. As a result, the computational complexity of the enumeration-based method to directly compute the discrete conditional interference cdf is exponential in the number of co-channel GBSs, which is prohibitive for implementation. To reduce such high complexity, we propose a new and more efficient method, named the \textit{lattice approximation (LA)} method.

\subsection{Downlink SNR Distribution}

For any given UAV location $\bold u$ and by dropping $(\bold u)$ for brevity, the SNR $\gamma_{\textrm{dl}}$ in \eqref{DLsinr} can be rewritten as
\begin{equation}\label{DLsinrRewritten}
\gamma_{\textrm{dl}}=\frac{C_{i_s}}{\alpha_0+I}=\frac{\max\limits_{i\in\mathcal{B}} \big(\delta_i C_{\textrm{L},i}+ (1-\delta_i) C_{\textrm{NL},i}\big)}{\alpha_0+\sum\limits_{i\in\mathcal{B}_{\textrm{f}_s}\setminus i_s} \mu_i\big(\delta_i C_{\textrm{L},i}+ (1-\delta_i) C_{\textrm{NL},i}\big)},
\end{equation}
which is mainly determined by the UAV-GBS association and the aggregate co-channel interference $I$. However, it is evident that they are coupled with each other. 
In the following, we first derive a general formula to obtain the downlink SNR cdf, by resolving the above coupling based on conditional probabilities. 

From \eqref{DLsinrRewritten}, the cdf of $\gamma_{\textrm{dl}}$ can be defined as
\begin{align}\label{cdfSINR}
F_{\gamma_{\textrm{dl}}}(y)&\triangleq \mathbb{P}\{\gamma_{\textrm{dl}}\leq y\}=\mathbb{P}\bigg\{\frac{C_{i_s}}{\alpha_0+I}\leq y\bigg\}\notag\\
&=\mathbb{P}\big\{I\geq -\alpha_0+C_{i_s} /y\big\}\notag\\
&\stackrel{(a)}{=}\mathbb{E}_{C_{i_s}}\bigg\{\mathbb{P}\big\{I\geq -\alpha_0+C_{i_s} /y|C_{i_s}\big\} \bigg\}\notag\\
&=\mathbb{E}_{C_{i_s}}\bigg\{1-F_{I|C_{i_s}} \big(-\alpha_0+C_{i_s} /y\big)\bigg\},
\end{align}
where $y>0$ is assumed; the right-hand side (RHS) of $(a)$ takes expectation of the conditional probability $\mathbb{P}\big\{I\geq -\alpha_0+C_{i_s} /y|C_{i_s}\big\}$ over the realization of the channel power gain $C_{i_s}$ with the associated GBS; and $F_{I|C_{i_s}}(\cdot)$ is the cdf of $I$ conditioned on the realization of $C_{i_s}$.

The realization of $C_{i_s}$ depends on the UAV-GBS association, whose pmf can be obtained similarly by Algorithm \ref{AlgUplink} in Section \ref{SectionUplink}.
The overall algorithm to obtain the cdf of the downlink SNR $\gamma_{\textrm{dl}}$ is summarized in Algorithm \ref{AlgDownlink}.
Denote $T$ as the running time of each iteration (which will be specified later in Section \ref{SectionDFTCF}) to obtain the conditional interference cdf $F_{I|C_{i_s}}(\cdot)$ in Step 4 of this algorithm. As a result, the worse-case complexity to compute the cdf of downlink SNR $\gamma_{\textrm{dl}}(\bold u)$ at UAV location $\bold u$ is $O(T|\mathcal{B}|)$.
Note that similar to Algorithm \ref{AlgUplink}, we can stop the algorithm early if the current LoS channel power gain $C_{\textrm{L},i_m}$ is smaller than $C_{\textrm{NL}}^{\textrm{max}}$, which is omitted in Algorithm \ref{AlgDownlink} for brevity. The early stopping threshold $\epsilon$ for Algorithm \ref{AlgUplink} can also be similarly used to further reduce the number of iterations.

\begin{algorithm}[H]\caption{Computing the cdf of downlink SNR $\gamma_{\textrm{dl}}$}\label{AlgDownlink}
\begin{small}
\begin{algorithmic}[1]
\STATE Sort $\{C_{\textrm{L},i}\}_{i\in\mathcal{B}}$ in descending order, and denote the ordered index of the GBSs as $i_m, m=1,\cdots,|\mathcal{B}|$.
\FOR{$m=1,\cdots,|\mathcal{B}|$}
\STATE Set $\mathbb{P}\{C_{i_s}=C_{\textrm{L},i_m}\}=p_{\textrm{L},i_m}\prod_{j=1}^{m-1}(1-p_{\textrm{L},i_j})$.
\STATE Obtain the conditional cdf $F_{I|C_{i_s}}(\cdot)$ by using the LA method in Section \ref{SectionDFTCF}.
\ENDFOR
\STATE Compute the cdf of $\gamma_{\textrm{dl}}$ by \eqref{cdfSINR}. 
\end{algorithmic}
\end{small}
\end{algorithm}

The main challenge of implementing Algorithm \ref{AlgDownlink} lies in how to obtain the conditional cdf $F_{I|C_{i_s}}(\cdot)$ in each iteration $m$, which is addressed in the following. 
For any given UAV location $\bold u$ and by dropping $(\bold u)$ for brevity, the aggregate interference power in \eqref{interferenceTotal} can be expressed as
\begin{equation}\label{interferenceTotalrv}
I=\sum\limits_{i\in\mathcal{B}_{\textrm{f}_s}\setminus i_s} \mu_i\big(\delta_i h_{\textrm{L},i}+ (1-\delta_i) h_{\textrm{NL},i}\big)G_{u,i}G_b\big(\theta_i\big),
\end{equation}
which depends on two sets of independent RVs, i.e., the random channel activity $\mu_i$ and the probabilistic LoS channel occurrence $\delta_i$ of all the co-channel GBSs $i\in\mathcal{B}_{\textrm{f}_s}\setminus i_s$.
Conditioned on the realization of channel power gain $C_{i_s}$ of the associated GBS $i_s$, a subset of LoS channel realizations $\delta_i,i\in\mathcal{B}$ are implied.
Specifically, for the $m$-th iteration in Algorithm \ref{AlgDownlink}, some of the co-channel GBSs should have the NLoS channel realization, i.e., $\delta_{i_j}=0, \forall j=1,\cdots,m-1$. As a result, the 
involved co-channel interference terms $I_{i_j}, j=1,\cdots,m-1, i_j\in\mathcal{B}_{\textrm{f}_s}\setminus i_s$, can only take \textit{two} possible values between 0 (corresponding to $\mu_i=0$) and $C_{\textrm{NL},i_j}$ (corresponding to $\mu_i=1$) with probabilities $1-\omega_{\textrm{dl},i_j}$ and $\omega_{\textrm{dl},i_j}$, respectively. On the other hand, the interference term $I_i$ from each remaining co-channel GBS $i\in\mathcal{B}_{\textrm{f}_s}\setminus i_s$ can take \textit{three} possible values of 0, $C_{\textrm{NL},i}$ and $C_{\textrm{L},i}$ with probabilities $1-\omega_{\textrm{dl},i}$, $\omega_{\textrm{dl},i}(1- p_{\textrm{L},i})$ and $\omega_{\textrm{dl},i} p_{\textrm{L},i}$, respectively.

In general, consider a discrete RV $\zeta_i$ which takes values from $1,\cdots,L$ with probabilities $p_{i,1},\cdots,p_{i,L}$, respectively, where $\sum_{l=1}^L p_{i,l}=1$. 
Based on independent $\zeta_i, i=1,\cdots,M$, a new RV is defined as
\begin{equation}\label{Z}
Z\triangleq\sum\limits_{i=1}^M z_i=\sum\limits_{i=1}^M\sum_{l=1}^L a_{i,l}\mathbf{1}_{l}(\zeta_i),
\end{equation}
which is named as the \textit{generalized Poisson multinomial (GPM)} RV\footnote{It 
is called ``generalized" as it allows each individual summand $z_i$ to take values from different real-valued sample spaces with different probabilities, thus extending the Poisson multinomial distribution\cite{PoissonMultinomial} and further the ordinary multinomial distribution.},
where $z_i\triangleq \sum_{l=1}^L a_{i,l}\mathbf{1}_{l}(\zeta_i)$ can take $L$ possible values $a_{i,1},\cdots,a_{i,L}$, and $\mathbf{1}_{l}(x)$ is the indicator function where $\mathbf{1}_{l}(x)=1$ when $x=l$ and $\mathbf{1}_{l}(x)=0$ otherwise. Without loss of generality, assume that $a_{i,1}\leq a_{i,2}\cdots\leq a_{i,L}$.
From \eqref{interferenceTotalrv} and \eqref{Z}, it follows that the conditional aggregate interference $I$ can be modeled as a GPM variable.

For real-valued $a_{i,l}$'s in practice, the value space of the composite GPM variable $Z$ has a worst-case size of $L^M$.
As a result, in order to completely characterize the pmf of $Z$, an enumeration-based method has time complexity $O(L^{M})$, and moreover it requires additional $O(L^{M})$ time for sorting the mass points in order to obtain the cdf.
To avoid such exponential complexity, in the sequel we propose the LA method to obtain the approximated cdf of $Z$ efficiently with high accuracy.

\subsection{Lattice Approximation (LA) Method}\label{SectionDFTCF}

Instead of directly computing the pmf/cdf, we leverage the characteristic function (cf)\footnote{The cf of an RV $X$ is defined as $\varphi_X(s)\triangleq \mathbb{E}\{e^{\mathbf{j}sX}\}=\int_{-\infty}^{\infty} e^{\mathbf{j}sx} \diff F_X(x)$, where $\mathbf{j}=\sqrt{-1}$.} of $Z$, which always exists and uniquely characterizes its distribution.
The cf approach is useful in the analysis of linear combinations of independent RVs. In our setting, the GPM variable $Z$ is the sum of independent RVs $z_i,i=1,\cdots, M$, whose cf is given by
\begin{equation}\label{cfAll}
\varphi_{Z}(s)\triangleq \mathbb{E}\{e^{\mathbf{j}sZ}\}=\prod\limits_{i=1}^M \varphi_{z_i}(s),
\end{equation}
which is decomposed as the product of individual cf $\varphi_{z_i}(s)$ given by
\begin{equation}\label{cfIndividual}
\varphi_{z_i}(s)\triangleq \mathbb{E}\{e^{\mathbf{j}s z_i}\}=\sum\limits_{l=1}^L p_{i,l}e^{\mathbf{j}s a_{i,l}}, i=1,\cdots, M.
\end{equation}
After decomposition, computing the cf $\varphi_{Z}(s)$ at one point $s$ has only linear complexity $O(LM)$, which involves computing $M$ independent cfs $\varphi_{z_i}(s),i=1,\cdots, M$, using the given values of $p_{i,l}$ and $a_{i,l}$ for $i=1,\cdots,M$ and $l=1,\cdots, L$.

Next, we introduce a cf inversion lemma which will be used to recover the cdf of $Z$.

\newtheorem{lemma}{Lemma}

\begin{lemma}\label{lemLattice}
If an RV $X$ has all its masses concentrated on the integer lattice $\mathcal{N}\triangleq\{0,1,\cdots,N-1\}$ of finite length $N$, then its pmf $q_n=\mathbb{P}\{X=n\}, n\in\mathcal{N}$ can be recovered by performing inverse discrete Fourier transform (DFT) on the $N$ equally-spaced samples of the cf $\varphi_X(s)$, i.e., 
\begin{equation}\label{IDFT}
q_n=\frac{1}{N}\sum\limits_{k=0}^{N-1} \varphi_X[k] e^{-\mathbf{j}2\pi kn/N}, n\in\mathcal{N},
\end{equation}
where $\varphi_X[k]\triangleq \varphi_X(s)|_{s=\frac{2\pi k}{N}}, k\in \mathcal{N}$.
\end{lemma}
\textit{Proof:} If $X$ has its masses concentrated on $\mathcal{N}$ with the pmf $q_n\triangleq\mathbb{P}\{X=n\}, n\in\mathcal{N}$, then its cf is given by
\begin{equation}
\varphi_X(s)\triangleq \mathbb{E}\{e^{\mathbf{j}sX}\}=\sum\limits_{n=0}^{N-1} q_n e^{\mathbf{j}ns},
\end{equation}
which is continuous in $s$ and has a period $2\pi$ (analogous to the frequency domain signal), and coincides with the discrete time Fourier transform (DTFT) for the ``time" sequence $q_n, n\in\mathcal{N}$.
To obtain perfect recovery (i.e., no aliasing) of the sequence $q_n, n\in\mathcal{N}$, we can take $N$ samples from $\varphi_X(s)$ which are equally spaced around the unit circle as
\begin{equation}
\varphi_X[k]\triangleq \varphi_X(s)|_{s=\frac{2\pi k}{N}}=\sum\limits_{n=0}^{N-1} q_n e^{\mathbf{j}2\pi kn/N}, k\in \mathcal{N},
\end{equation}
which is the DFT of the sequence $q_n, n\in\mathcal{N}$.
Therefore, we can use the inverse DFT to recover the pmf sequence as in \eqref{IDFT} and hence lemma \ref{lemLattice} follows.
$\blacksquare$

Note that the fast Fourier transform (FFT) algorithm can be applied to perform the inverse DFT efficiently.
In our context, the cf samples for the GPM variable $Z$ can be obtained using \eqref{cfAll}.
However, Lemma \ref{lemLattice} cannot be applied directly to recover the pmf of $Z$.
This is because the real-valued discrete GPM variable $Z$ has its masses concentrated on a set of $L^M$ points, which are usually not equally spaced to form a lattice\footnote{A lattice is a set of the form $\{nd+\kappa| n=0,\pm 1, \pm 2, \cdots, \textrm{for some $d>0$ and $\kappa\in\mathbb{R}$}\}$.} or even integer lattice, and moreover the number of mass points to be recovered is overwhelming even for the FFT algorithm.
To resolve the above challenge, we propose the LA method, where the original $Z$ is approximated by an integer lattice with a bounded size of value space, and furthermore the efficient FFT algorithm is applied on its cf samples to recover its pmf and hence the approximated cdf of $Z$.

Specifically, we propose the \textit{offset-scale-quantize} operation to convert a general non-lattice GPM variable $Z$ into an integer-lattice GPM variable $\tilde{Z}$, and then use the cdf of $\tilde{Z}$ to obtain an approximated cdf of $Z$.
Denote $A_0\triangleq\sum_{i=1}^M a_{i,1}$ as the minimum possible value of $Z$, and $A\triangleq\sum_{i=1}^M a_{i,L}-A_0$ as the absolute range of possible values of $Z$.
The first step is to offset each summand $z_i$ by $a_{i,1}$, respectively, so that each summand has a minimum value of 0 and so does their sum.
This offsetting process is practically helpful for the stability of numerical computations involving complex numbers of potentially large phase (e.g., computing the cf samples in \eqref{cfAll}), and also facilitates the subsequent scaling operation.
Second, we perform a scale-and-quantize operation on $z_i-a_{i,1}$ and convert it into $\tilde{z}_i$, which is obtained by multiplying $z_i-a_{i,1}$ with a common factor of $\beta>0$, and then rounding all the possible values $\beta (a_{i,l}-a_{i,1})$'s to their nearest integer values, denoted as $\tilde a_{i,l}$, respectively, to which the original mass $p_{i,l}$ now attributes\footnote{Rounding is performed on the individual summand instead of the sum, since otherwise we still have to round $L^M$ possible values of the sum.}.
The absolute value range of the converted GPM variable $\tilde{Z}\triangleq \sum_i^M \tilde{z}_i$ can then be denoted as
\begin{equation}
\tilde A\triangleq \sum_{i=1}^M \tilde a_{i,L},
\end{equation}
which is around $\lceil \beta A\rceil$ with $\lceil \cdot\rceil$ denoting the ceiling function.
As a result, the new GPM variable $\tilde{Z}$ has its masses concentrated on the lattice $\mathcal{N}=\{0,1,\cdots,\tilde A\}$ of finite length $\tilde A+1\approx\lceil \beta A\rceil +1$.
Compared to the exponential number $L^M$ of mass points in a general non-lattice GPM variable $Z$, the scale-and-quantize operation effectively merges adjacent non-lattice mass points into an integer lattice  with bounded size of value space, where the size can be controlled by the scaling factor $\beta$.
Note that the offset-scale-quantize operation takes effect on each possible value of each summand $z_i$, which has an overall complexity of $O(LM)$.

For the converted integer-valued GPM variable $\tilde{Z}$, samples of its cf $\varphi_{\tilde{Z}}(s)$ can be obtained similar to \eqref{cfAll}.
Therefore, we can apply Lemma \ref{lemLattice} to obtain the exact pmf of $\tilde{Z}$ whose mass points are already sorted, and hence we can directly obtain its cdf $F_{\tilde{Z}}(\cdot)$.
The cdf of the original GPM variable $Z$ can then be approximated as
\begin{align}\label{cdfReal}
F_Z(x)&\triangleq \mathbb{P}\{Z\leq x\}=\mathbb{P}\{\beta(Z-A_0)\leq \beta(x-A_0)\}\notag\\
&\stackrel{(a)}{\approx} \mathbb{P}\{\tilde{Z}\leq \beta(x-A_0)\}= F_{\tilde{Z}}\big(\beta(x-A_0)\big),
\end{align}
where $(a)$ is due to possible quantization errors.

Note that a larger scaling factor $\beta$ corresponds to more lattice points and hence smaller quantization errors. However, the time to compute $\beta A$ samples of the cf $\varphi_{\tilde{Z}}(s)$ in \eqref{cfAll} increases in $O(\beta ALM)$, while the time to perform FFT increases in $O\big( \beta A\log_2 ( \beta A)\big)$.
Therefore, there is a trade-off between the accuracy and computational efficiency.
The overall algorithm to obtain the cdf of a real-valued GPM variable is summarized in Algorithm \ref{AlgDFTcf}, whose computational time $T$ has the complexity of $O(LM)+O(\beta A LM)+O\big(\beta A\log_2 ( \beta A)\big)$.
Note that $\beta$ can be chosen as $\beta=c_0/A$ so that $\beta A$ is a large enough constant $c_0$ (e.g., in the range of 100 to 1000) to provide high approximation accuracy, and hence the LA method has a linear complexity in $M$.

\begin{algorithm}[H]\caption{LA method to approximate the cdf of non-lattice GPM}\label{AlgDFTcf}
\begin{small}
\textbf{Input:} For each independent summand $z_i, i=1,\cdots, M$, given its possible values $a_{i,l}$ with probabilities $p_{i,l}$ for $l=1,\cdots,L$. Let $a_{i,1}\leq a_{i,2}\cdots\leq a_{i,L}$.\\
\textbf{Output:} The approximated cdf of the GPM variable $Z=\sum_{i=1}^M z_i$.
\begin{algorithmic}[1]
\STATE $A_0\triangleq\sum_{i=1}^M a_{i,1}; A\triangleq\sum_{i=1}^M a_{i,L}-A_0$.
\STATE Set $\beta=c_0/A$.
\STATE Under $\beta$, perform the offset-scale-quantize operation on $z_i$ and convert it into $\tilde{z}_i$, $i=1,\cdots, M$.
\STATE Let $\tilde{Z}\triangleq \sum_i^M \tilde{z}_i$. Apply Lemma \ref{lemLattice} to obtain its sorted pmf and hence its cdf $F_{\tilde{Z}}(\cdot)$.
\STATE Approximate the cdf of $Z$ based on \eqref{cdfReal}.
\end{algorithmic}
\end{small}
\end{algorithm}

\section{Numerical Results}\label{SectionSimulation}



Consider a network area centered at the reference GBS 0 with radius $D_\textrm{max}$. 
The following parameters are used if not mentioned otherwise: $D_\textrm{max}=10D$, $D=500$ m, $H_b=20$ m, $\theta_\textrm{tilt}=-10^\circ$, $\Phi_u=90^\circ$ (isotropic UAV antenna), $f_c=2$ GHz, $c=3\times 10^8$ m/s, per-RB noise power $\sigma^2=-124$ dBm, $P_b=0.1$ W, $P_u=-20$ dBm\footnote{$P_u$ is set to a low level (40 dB lower than $P_b$) in order to limit the uplink interference to other cells.}, $\epsilon=10^{-6}$, $c_0=1000$, $\eta_\textrm{ul}=12$ dB, $\eta_\textrm{dl}=2$ dB and $\omega_\textrm{dl}=0.5$. The corresponding empirical formulas in \cite{3GPPevaluationAssumption} are used in the simulation for the channel model in \eqref{probLOS}.

\subsection{Uplink Communication}

\subsubsection{Spatial Distribution of Non-Outage Probability}

\begin{figure}
        \centering
        \begin{subfigure}[b]{0.65\linewidth}
                \includegraphics[width=1\linewidth,  trim=0 0 0 0,clip]{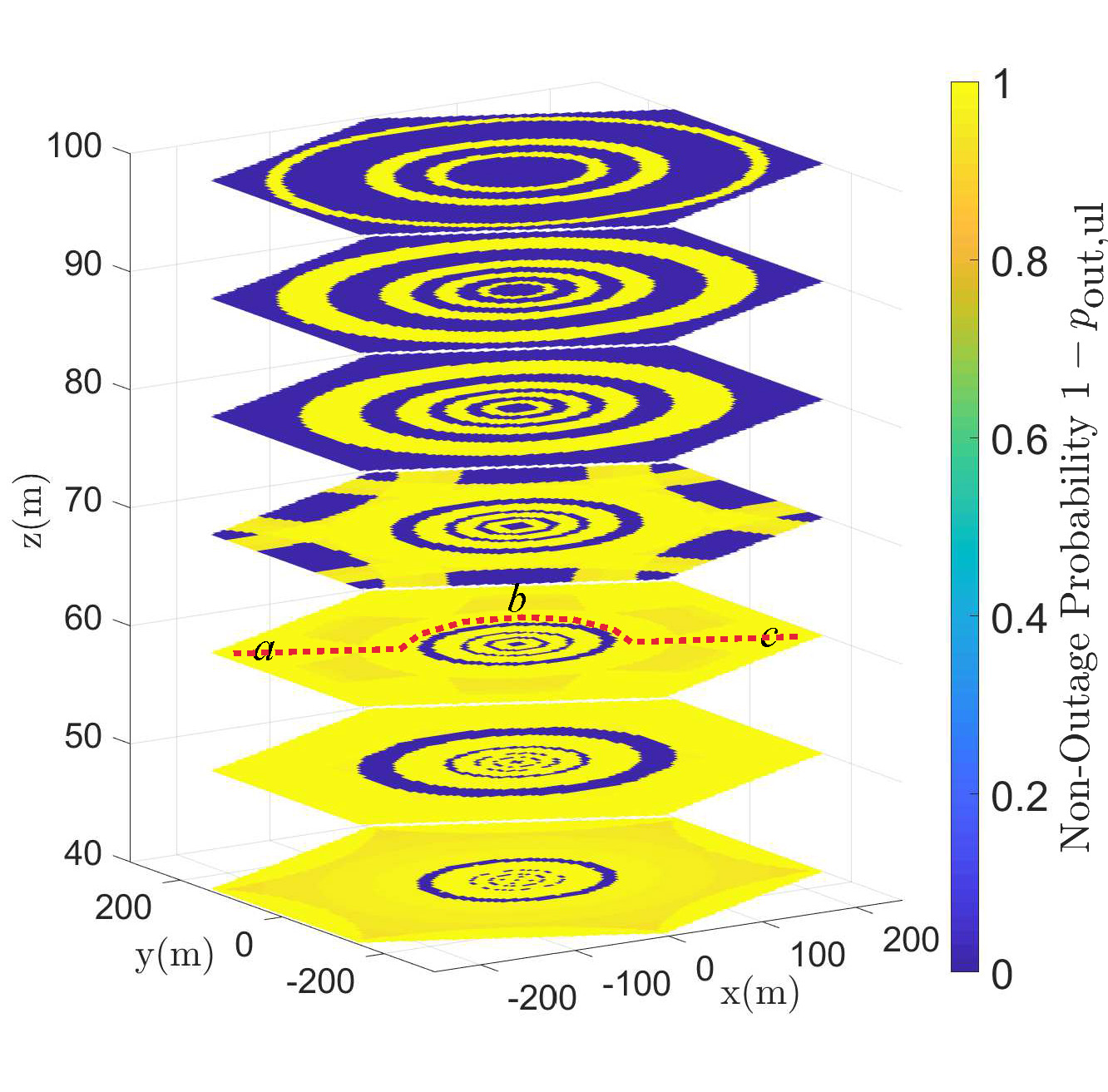}
                \caption{Under $\theta_\textrm{tilt}=-10^\circ$.}
                \label{PoutCell0Uplink75}
        \end{subfigure}%
                 
        \begin{subfigure}[b]{0.65\linewidth}
                \includegraphics[width=1\linewidth,  trim=0 0 0 0,clip]{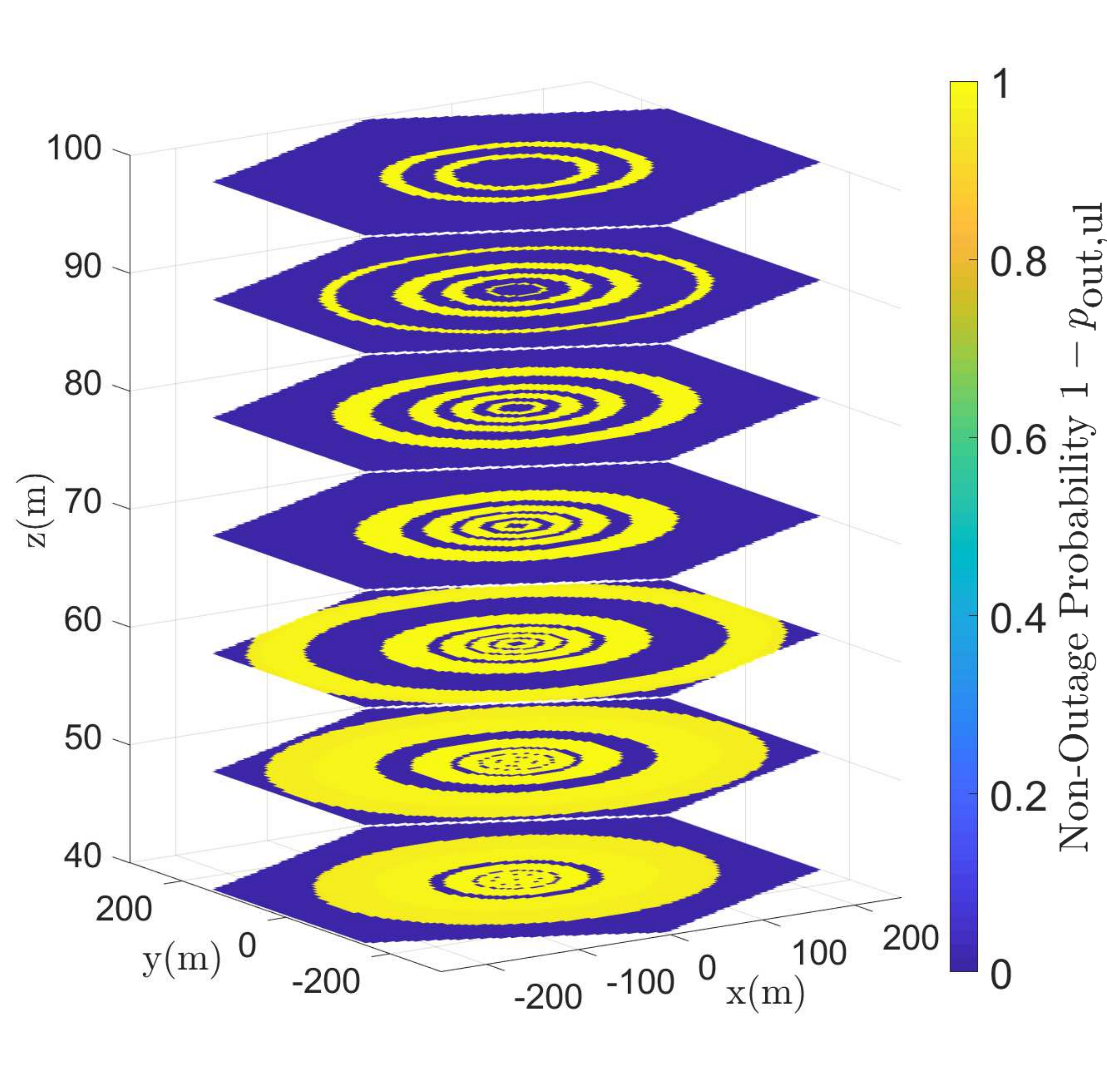}
                \caption{Under $\theta_\textrm{tilt}=-20^\circ$.}
                \label{PoutCell0Uplink150}
        \end{subfigure}%
        \caption{Uplink non-outage probability $1-p_\textrm{out,ul}$ for the UAV located in the reference cell 0 at different altitude $H_u$.}\label{PoutCell0Uplink}
\end{figure}

The non-outage probability $1-p_\textrm{out,ul}(\bold u)$ at location $\bold u$ can be obtained by the analysis in Section \ref{SectionUplink}.
The spatial distribution of uplink non-outage probability for the UAV located in the reference cell 0 at different altitude $H_u$ under different GBS antenna downtilt angle $\theta_\textrm{tilt}$ (as in Fig. \ref{BSantennaPlot} (b) and (c)) is plotted in Fig. \ref{PoutCell0Uplink}.
It can be seen that the GBS antenna pattern, especially the sidelobes and nulls in between, has a significant impact on the spatial distribution of non-outage probability, where the UAV is mainly served by the sidelobe peaks of GBS 0 or other GBSs complementarily. 
Moreover, a larger $|\theta_\textrm{tilt}|$ leads to overall smaller GBS antenna gain for the UAV above the GBS height, which results in overall lower strength of the communication link and hence lower coverage probability (average non-outage probability), as will be also shown later in Fig. \ref{ULpcHcurve}.
On the other hand, as the UAV altitude increases, the UAV-GBS link distance increases, while so does the LoS probability. This two factors affect the link strength and hence the coverage probability in opposite ways.

The spatial distribution of non-outage probability is helpful in facilitating UAV path planning/movement control.
For the example in Fig. \ref{PoutCell0Uplink}(a), in order to move across the cell from location $a$ to location $c$, the straight path $a-c$ needs to fly over the outage region while the dashed red trajectory $a-b-c$ enjoys full uplink coverage along the way with short traveling distance. Other examples may suggest UAV movement across different layers of altitude in the 3D space. Similar applications exist for the downlink case (e.g., based on Fig. \ref{PoutDL90} in the sequel).

\subsubsection{Coverage Probability}

\begin{figure}
\centering
   \includegraphics[width=0.95\linewidth]{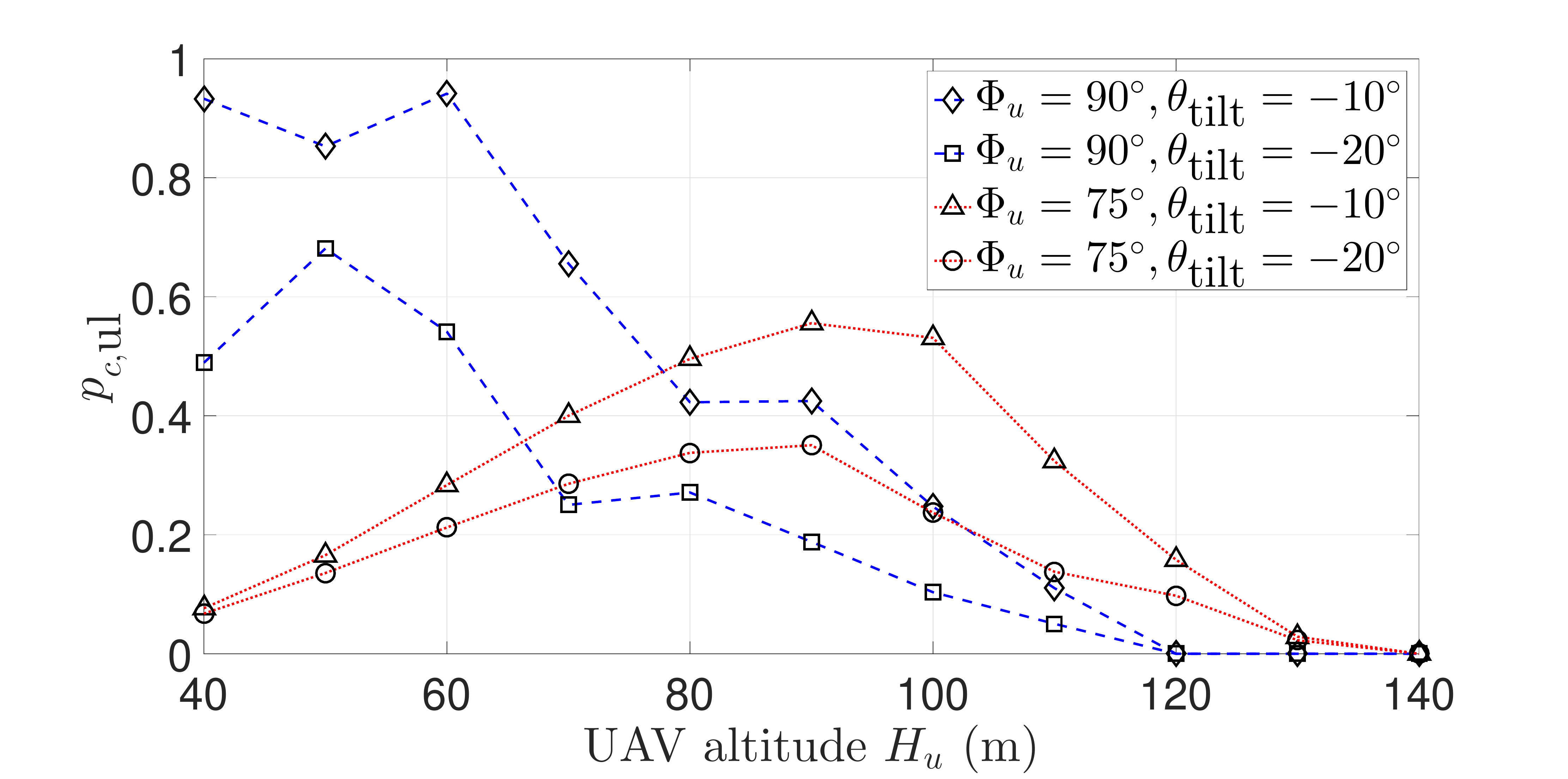}
\caption{Uplink coverage probability versus the UAV altitude.\vspace{-2ex}}\label{ULpcHcurve}
\end{figure}

The uplink coverage probability $p_\textrm{c,ul}$ at an altitude $H_u$ can be obtained in \eqref{PcUL2d}.
The trends of $p_\textrm{c,ul}$ versus the UAV altitude $H_u$ under different GBS antenna downtilt angle $\theta_\textrm{tilt}$ and UAV antenna beamwidth $\Phi_u$ are plotted in Fig. \ref{ULpcHcurve}.
For the cases with $\Phi_u=90^\circ$ (isotropic UAV antenna), the coverage probability $p_\textrm{c,ul}$ corresponds to the spatial average of non-outage probability in Fig. \ref{PoutCell0Uplink} and tends to decrease with the UAV altitude $H_u$, while the oscillation is mainly due to the non-uniform GBS antenna pattern in the elevation domain.
In contrast, for the cases with $\Phi_u=75^\circ$, the UAV directional antenna gain is helpful for $p_\textrm{c,ul}$ (see Fig. \ref{ULpcHcurve} at high altitude, e.g., $H_u\geq 80$ m). On the other hand, at low altitude, the limited coverage range of the UAV antenna mainlobe limits its chance to be served by GBSs with good channel. As a result, the coverage probability first increases and then decreases with the UAV altitude. Moreover, the curve also becomes smoother since the UAV is potentially served by fewer GBSs with non-uniform antenna patterns.

\begin{figure}
\centering
   \includegraphics[width=0.95\linewidth]{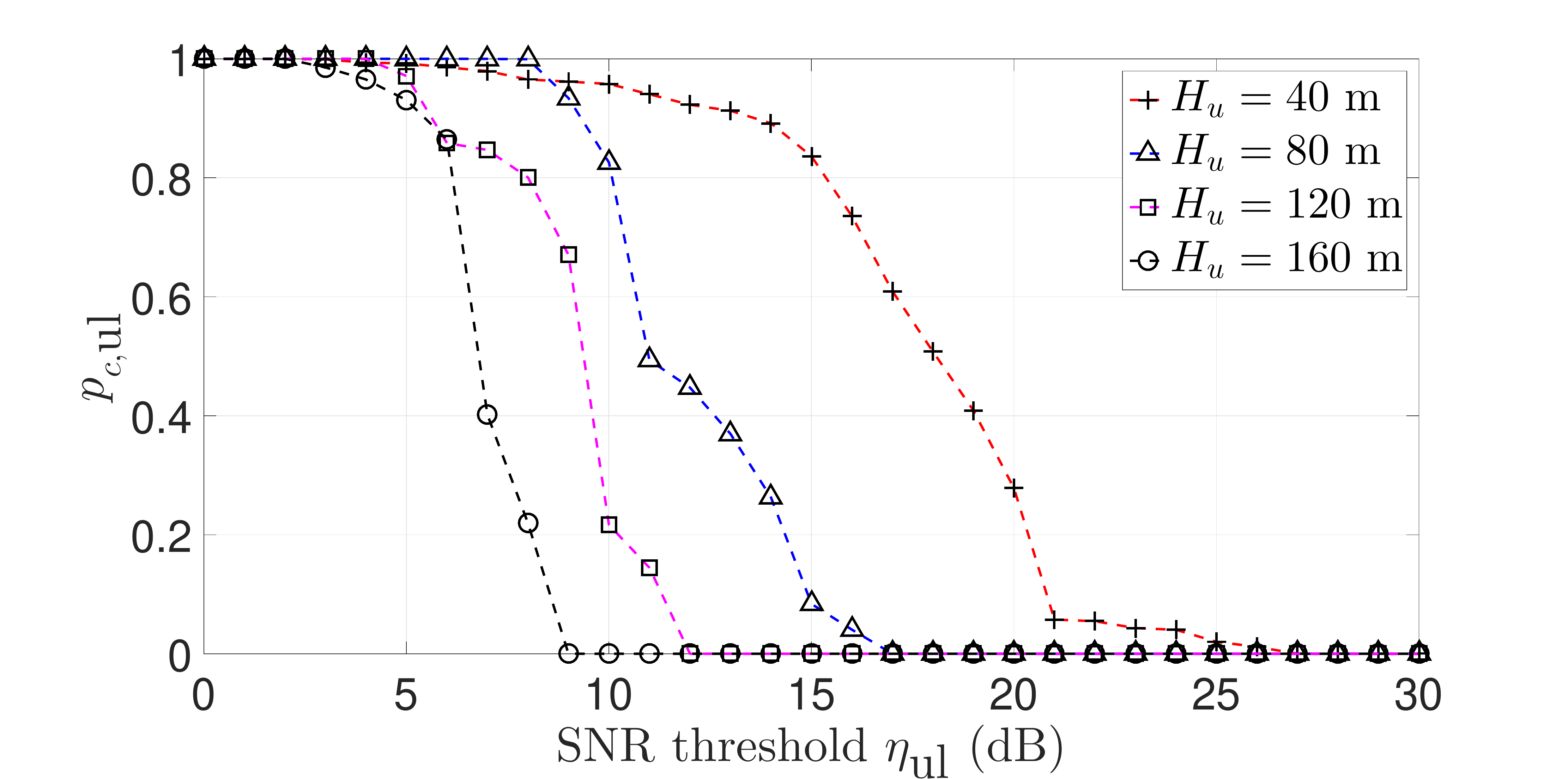}
\caption{Uplink coverage probability $p_{c,\textrm{ul}}$ versus the SNR threshold $\eta_\textrm{ul}$, under $\Phi_u=90^\circ$ (isotropic UAV antenna) and GBS antenna downtilt $\theta_\textrm{tilt}=-10^\circ$. \vspace{-2ex}}\label{ULpcETAcurve}
\end{figure}

In addition, the trend of $p_{c,\textrm{ul}}$ versus the SNR threshold $\eta_\textrm{ul}$ is plotted in Fig. \ref{ULpcETAcurve}.
It can be seen that $p_{c,\textrm{ul}}$ in general decreases with $\eta_\textrm{ul}$ at a given UAV altitude.
On the other hand, as the UAV altitude $H_u$ increases, the LoS probability also increases, resulting in an overall less variation of $p_{c,\textrm{ul}}$ versus $\eta_\textrm{ul}$.

\subsection{Downlink Communication}

\subsubsection{Aggregate Interference Distribution}\label{SectionSimulationI}

Assume that the UAV is equipped with isotropic antenna of unit gain, located at horizontal location $(0.3D,0.1D)$ with altitude $H_u=100$ m, and is associated with the GBS with the largest LoS channel gain.
We apply the LA method, enumeration method, Monte Carlo (MC) based simulation, and a benchmark Gaussian Approximation (GA) method to obtain the aggregate interference distribution under different cell loading factor $\omega_\textrm{dl}$, for comparison.
For the MC method, $10^6$ random samples of the aggregate interference $I$ is generated in order to provide a good approximation for the true distribution, where each sample of $I$ is drawn by summing over one realization of the interference terms $I_i, i\in\mathcal{B}_{\textrm{f}_s}\setminus i_s$ which are randomly and independently generated.
On the other hand, the GA method is based on the central limit theorem, where the cdf of $I$ is approximated by the non-negative part of the Gaussian cdf which is normalized such that the total probability is 1.
The mean and standard deviation of $I$ need to be computed in the GA method, based on the given values of $a_{i,l}$'s and $p_{i,l}$'s and with a computation time $T$ of $O(LM^2)$.

\begin{figure}
\centering
   \includegraphics[width=1\linewidth]{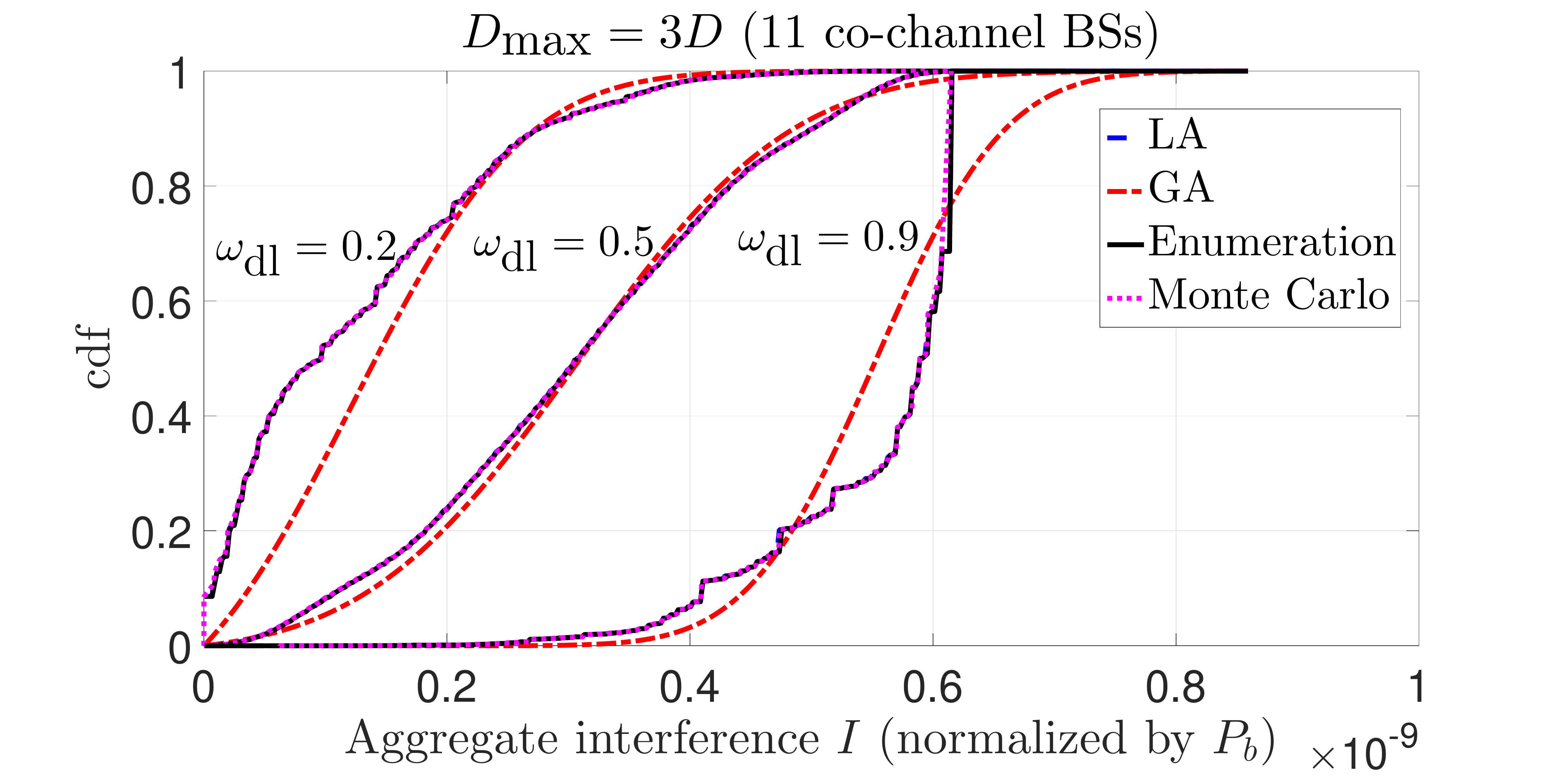}
\caption{The cdf of the downlink aggregate interference $I$ for network layout $D_\textrm{max}=3D$, under different cell loading factor $\omega_\textrm{dl}$.\vspace{-2ex}}\label{H100Dmax3}
\end{figure}

Consider a network area with radius $D_\textrm{max}=3D$, where there are 37 GBSs in $\mathcal{B}$ and 11 co-channel GBSs. The results are plotted in Fig. \ref{H100Dmax3}. 
It can be seen that the LA method matches almost exactly with the enumeration method, while both results are verified by MC simulation. 
On the other hand, the GA method provides fair approximation result for the case with moderate loading (e.g., $\omega_\textrm{dl}=0.5$), while the approximation result is poor for the case with low (or high) loading.
The average CPU time for the LA, GA, MC, and enumeration methods are 0.040, 0.028, 7.36 and 194 seconds, respectively, which are performed in MATLAB2015b on a laptop computer with Intel i7 2.7GHz CPU and 8GB memory without multi-core tasking.
It can be seen that both the LA and GA methods run much faster than the MC and enumeration methods.
Note that due to exponential time complexity, the enumeration method cannot be used for a larger setup with one or more tiers of co-channel GBSs (i.e., $M\geq 17$).
In summary, the LA method provides highly accurate approximation for the aggregate interference distribution with low time complexity, thus it will be applied in the rest of simulations.

\subsubsection{Spatial Distribution of Non-Outage Probability}

\begin{figure}
        \centering
        \begin{subfigure}[b]{0.65\linewidth}
                \includegraphics[width=1\linewidth,  trim=0 0 0 0,clip]{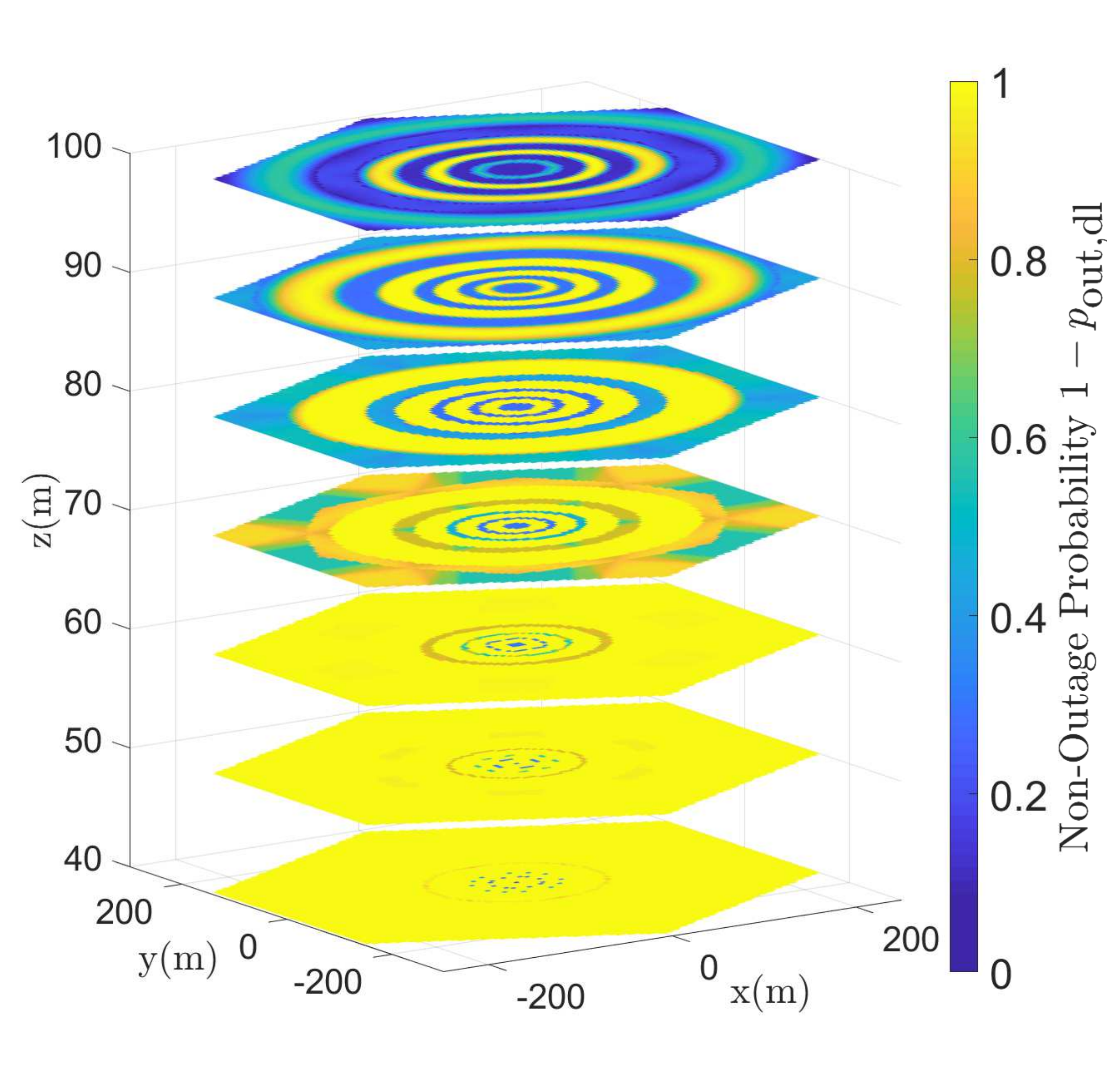}
                \caption{Under $\theta_\textrm{tilt}=-10^\circ$.}
                \label{PoutCell0DL75}
        \end{subfigure}%
                  
        \begin{subfigure}[b]{0.65\linewidth}
                \includegraphics[width=1\linewidth,  trim=0 0 0 0,clip]{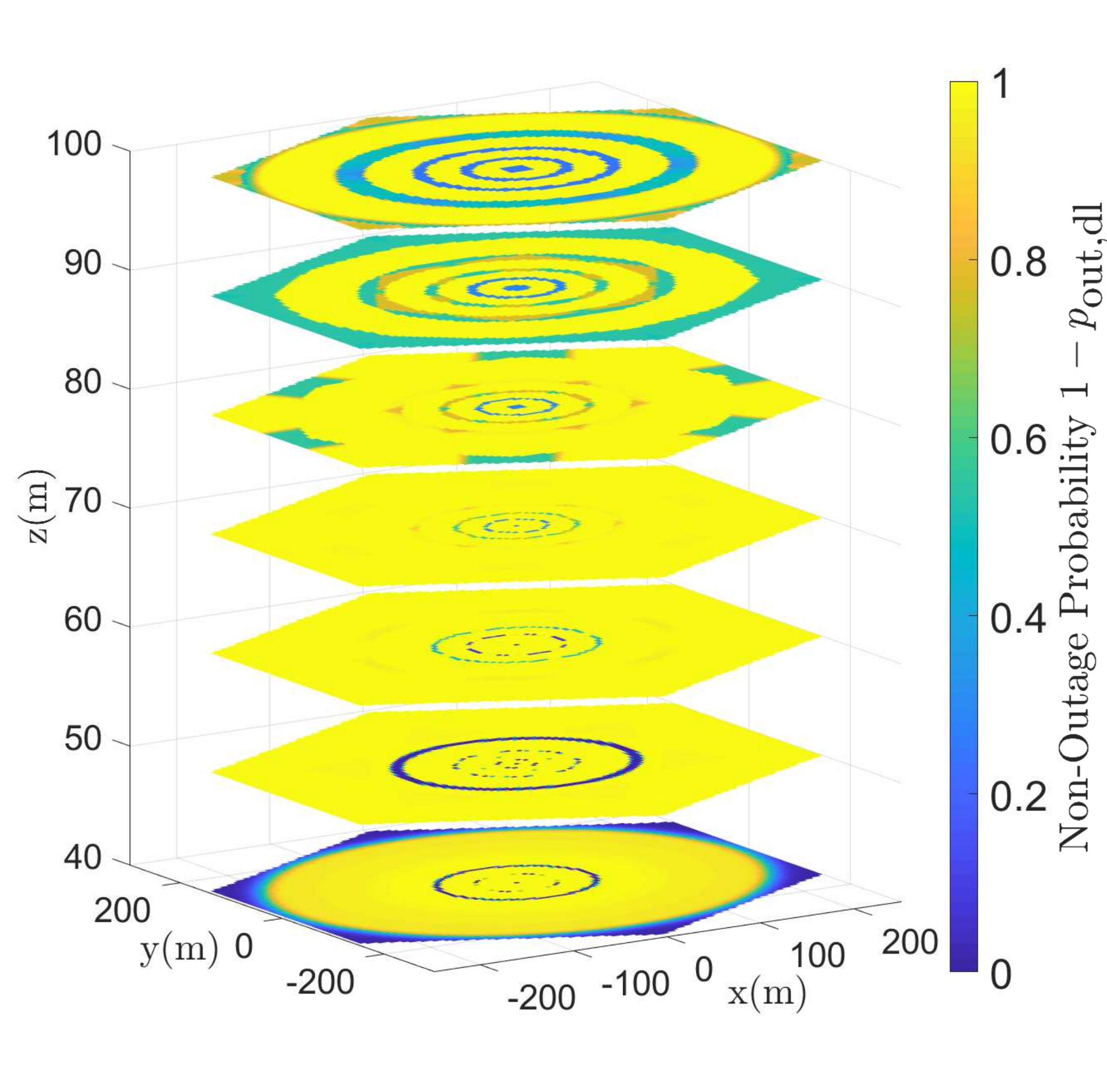}
                \caption{Under $\theta_\textrm{tilt}=-20^\circ$.}
                \label{PoutCell0DL150}
        \end{subfigure}%
        \caption{Downlink non-outage probability $1-p_\textrm{out,dl}$ for the UAV located in the reference cell 0 at different altitude $H_u$.}\label{PoutDL90}
\end{figure}

The downlink non-outage probability $1-p_\textrm{out,dl}(\bold u)$ at location $\bold u$ can be obtained by the analysis in Section \ref{SectionDownlinkMain}.
The spatial distribution of downlink non-outage probability for the UAV located in the reference cell 0 at different altitude $H_u$ is plotted in Fig. \ref{PoutDL90}.
Compared to the uplink, the UAV's non-outage probability in the downlink is determined by the direct communication link as well as the aggregate interference distribution.
In particular, a larger downtilt angle $|\theta_\textrm{tilt}|$ leads to overall smaller GBS antenna gain for the UAV above the GBS height, which reduces the overall strength of both direct link and interference links. However, the reduction on the aggregate interference is more significant, resulting in overall higher downlink coverage probability for the UAV above a certain altitude (e.g., $H_u\geq 60$ m), which is also shown next in Fig. \ref{DLpcHcurve}.

\subsubsection{Coverage Probability}

\begin{figure}
\centering
   \includegraphics[width=0.95\linewidth]{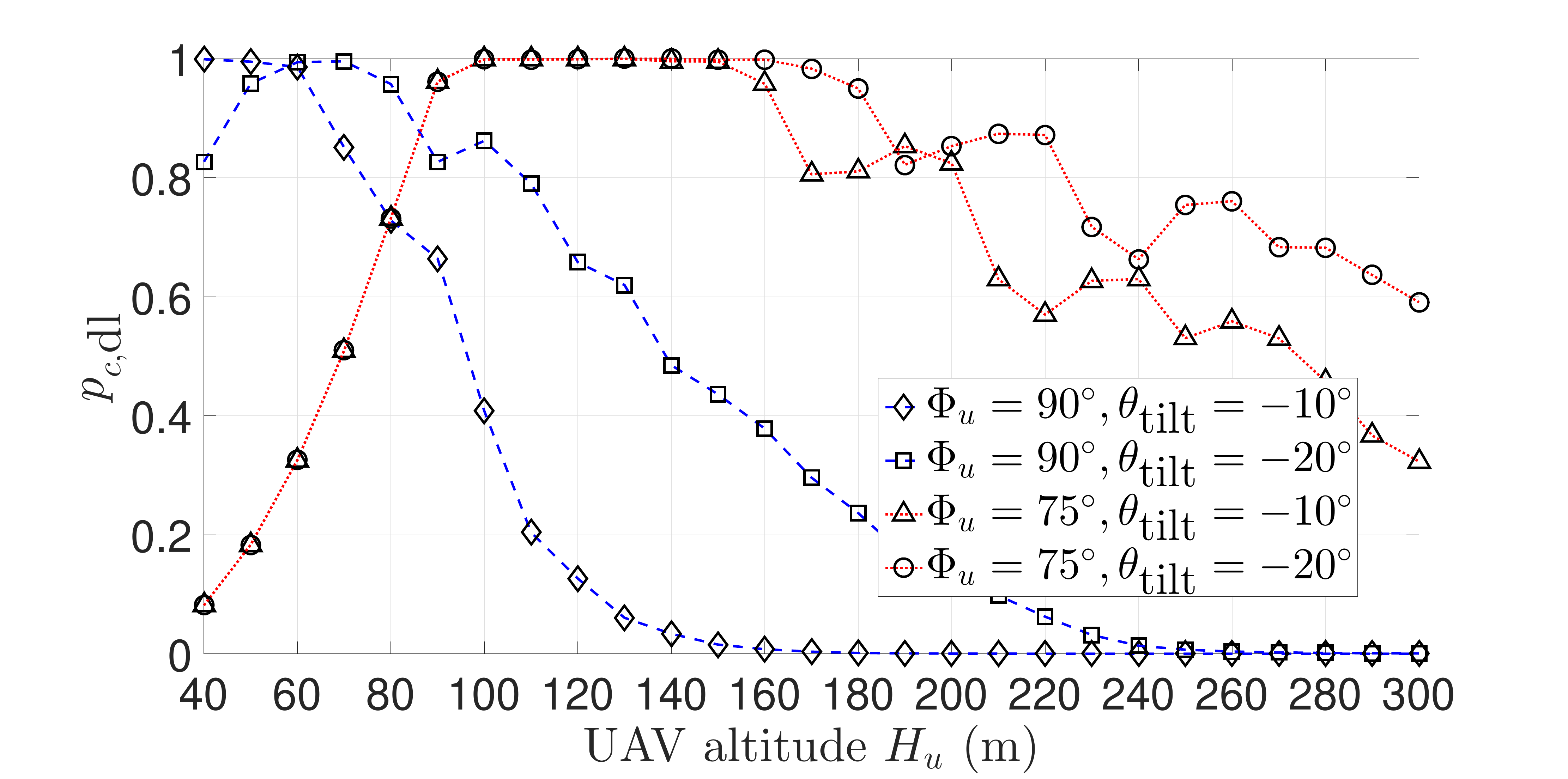}
\caption{Downlink coverage probability versus the UAV altitude.\vspace{-2ex}}\label{DLpcHcurve}
\end{figure}

The trends of downlink coverage probability $p_\textrm{c,dl}$ versus the UAV altitude $H_u$ under different UAV antenna beamwidth $\Phi_u$ and GBS antenna downtilt angle $\theta_\textrm{tilt}$ are plotted in Fig. \ref{DLpcHcurve}.
The curve oscillation is due to the non-uniform GBS antenna pattern which also results in complicated interference distribution.
It can be seen that at high altitude, the directional UAV antenna is helpful for improving $p_\textrm{c,dl}$, which not only brings directional antenna gain, but also effectively limits the interference from GBSs outside the UAV's antenna coverage range, hence resulting in higher $p_\textrm{c,dl}$ compared to the case with isotropic UAV antenna.

\begin{figure}
\centering
   \includegraphics[width=0.95\linewidth]{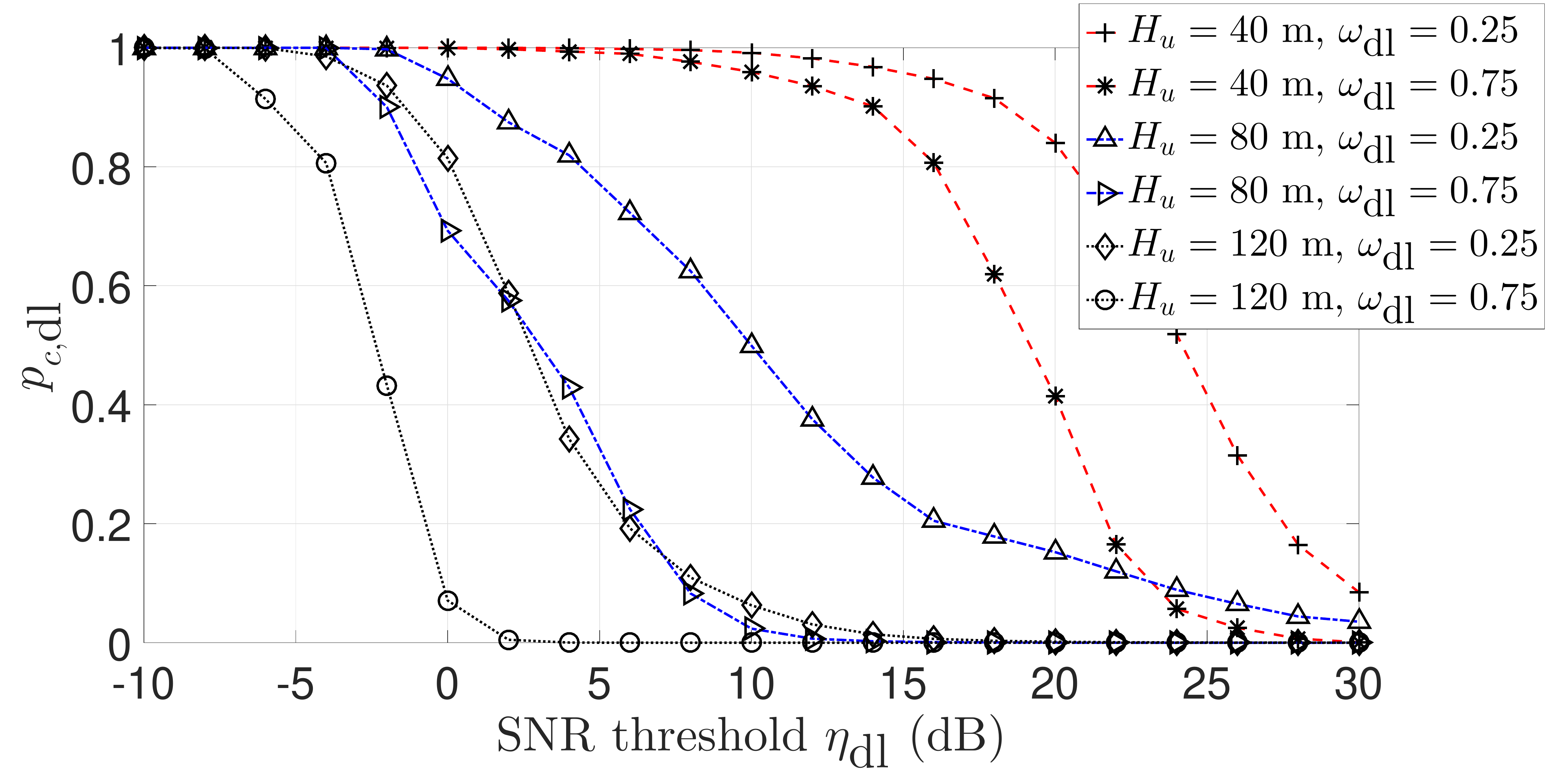}
\caption{Downlink coverage probability $p_{c,\textrm{ul}}$ versus the SNR threshold $\eta_\textrm{dl}$, under $\Phi_u=90^\circ$ (isotropic UAV antenna) and GBS antenna downtilt $\theta_\textrm{tilt}=-10^\circ$.\vspace{-2ex}}\label{DLpcETAcurve}
\end{figure}

In addition, the trend of $p_{c,\textrm{dl}}$ versus the SNR threshold $\eta_\textrm{dl}$ is plotted in Fig. \ref{DLpcETAcurve}.
It can be seen that at a given altitude, $p_{c,\textrm{dl}}$ decreases with $\eta_\textrm{dl}$, and is lower for a higher loading factor $\omega_\textrm{dl}$.
Moreover, a higher loading factor $\omega_\textrm{dl}$ also leads to overall less variation of $p_{c,\textrm{dl}}$ versus $\eta_\textrm{dl}$.

\section{Conclusions}\label{SectionConclusion}

This paper studies the 3D system modeling and coverage performance analysis for network-connected UAVs in the cellular uplink and downlink communications.
A 3D cellular network model is presented which incorporates the 3D air-ground channel and 3D patterns of the GBS/UAV antennas.
Based on it, an analytical framework is further proposed to characterize the uplink/downlink 3D coverage performance, which effectively reduces the exponential complexity due to UAV-GBS association and coupled interference distribution.
The conditional discrete interference is modeled as a new GPM RV, and a novel LA method is proposed to obtain the interference distribution efficiently with high accuracy.
The 3D coverage analysis is validated by extensive numerical results, which also show the effects of cell loading factor, GBS antenna downtilt, UAV altitude and antenna beamwidth, and applications for UAV path planning/movement control.

Our analytical framework is applicable to heterogeneous GBS locations, heights, antenna patterns and loading factors.
The results based on directional UAV antenna and practically downtilted GBS antenna provide motivation for more advanced antenna and beamforming design, such as vertically sectorized GBS antenna and 3D digital beamforming at the GBS/UAV.
Advanced interference mitigation techniques such as multi-cell coordinated GBS selection, channel allocation, power control and joint transmission/reception can also be applied to further improve the coverage performance, which will lead to promising future studies.

\bibliography{IEEEabrv,BibDIRP}

\end{document}